\documentclass{ieeeaccess}
\usepackage[hidelinks]{hyperref}
\usepackage{cite}
\usepackage{amsmath,amssymb,amsfonts}
\usepackage{graphicx}
\usepackage{textcomp}
\usepackage{latexsym}
\usepackage{mathtools}
\usepackage[ruled,vlined]{algorithm2e}
\usepackage{amsthm}
\theoremstyle{definition}
\newtheorem{exmp}{Example}
\newtheorem{rmrk}{Remark}
\usepackage{setspace}
\usepackage{siunitx}
\DeclareMathOperator*{\minimize}{minimize}
\begin{document}
\history{Date of publication xxxx 00, 2023, date of current version xxxx 00, 2023.}
\doi{xx.xxxx/ACCESS.xxxx.xxxxxxx}

\title{Convex Estimation of Sparse-Smooth Power Spectral Densities from Mixtures of Realizations with Application to Weather Radar}
\author{\uppercase{Hiroki Kuroda}\authorrefmark{1}, \IEEEmembership{Member, IEEE},
\uppercase{Daichi Kitahara}\authorrefmark{2}, \IEEEmembership{Member, IEEE},
\uppercase{Eiichi Yoshikawa}\authorrefmark{3, 4}, \IEEEmembership{Member, IEEE},
\uppercase{Hiroshi Kikuchi}\authorrefmark{5}, \IEEEmembership{Member, IEEE},
and \uppercase{Tomoo Ushio}\authorrefmark{2}, \IEEEmembership{Senior Member, IEEE}}

\address[1]{Department of Information and Management Systems Engineering, Nagaoka University of Technology, Niigata 940-2188, Japan}
\address[2]{Division of Electrical, Electronic and Infocommunications Engineering, Osaka University, Osaka 565-0871, Japan}
\address[3]{Aeronautical Technology Directorate, Japan Aerospace Exploration Agency, Tokyo 181-0015, Japan}
\address[4]{Department of Electrical and Computer Engineering, Colorado State University, Fort Collins 80521 CO, USA}
\address[5]{Center for Space Science and Radio Engineering, The University of Electro-Communications, Tokyo 182-8585, Japan}
\tfootnote{This work was supported in part by the Japan Society for the Promotion of Science (JSPS) KAKENHI under Grant JP21K17827 and JP21H01592.}

\markboth
{Kuroda \headeretal: Convex Estimation of Sparse-Smooth PSDs from Mixtures of Realizations with Application to Weather Radar}
{Kuroda \headeretal: Convex Estimation of Sparse-Smooth PSDs from Mixtures of Realizations with Application to Weather Radar}

\corresp{Corresponding author: Hiroki Kuroda (e-mail: kuroda@vos.nagaokaut.ac.jp).}

\begin{abstract}
In this paper, we propose a convex optimization-based estimation of sparse and smooth power spectral densities (PSDs) of complex-valued random processes from mixtures of realizations. While the PSDs are related to the magnitude of the frequency components of the realizations, it has been a major challenge to exploit the smoothness of the PSDs, because penalizing the difference of the magnitude of the frequency components results in a nonconvex optimization problem that is difficult to solve. To address this challenge, we design the proposed model that jointly estimates the complex-valued frequency components and the nonnegative PSDs, which are respectively regularized to be sparse and sparse-smooth. By penalizing the difference of the nonnegative variable that estimates the PSDs, the proposed model can enhance the smoothness of the PSDs via convex optimization. Numerical experiments on the phased array weather radar, an advanced weather radar system, demonstrate that the proposed model achieves superior estimation accuracy compared to existing sparse estimation models, regardless of whether they are combined with a smoothing technique as a post-processing step or not.
\end{abstract}


%\titlepgskip=-15pt

\maketitle

\section{Introduction}
\label{sect:introduction}
\PARstart{P}{ower} spectral density (PSD) of a random process
describes how power of the random process is distributed over frequency.
Estimation of the PSD from realizations of a random process is
a fundamental problem in science and engineering \cite{Stoica:BookPSD,Brockwell:BookPSD,Oppenheim:BookDTSP}.
For weather radar applications,
the PSD estimation is essential for the analysis of weather phenomena,
because the PSD contains information pertaining to the precipitation intensity and the Doppler velocity distribution \cite{Bringi:BookDopplerRadar,Zrnic:AnalysisPSD,Janssen:AnalysisPSD,Wang:AnalsisPSD,Yu:AnalysisPSD}.
For example, the parabolic Doppler weather radar \cite{Bringi:BookDopplerRadar}
transmits a pencil beam and subsequently observes backscattered signals in a narrow range of elevation angles, which can be regarded as realizations of a single random process whose PSD reflects the weather condition in the narrow range.

We consider the estimation of PSDs from mixtures of realizations of random processes,
which is much more challenging than
the classical case of a single random process.
Our primary interest is on
the phased array weather radar (PAWR)
\cite{Zrnic:PAWR,McLaughlin:PAWR,Isom:PAWR,Yoshikawa:PAWR,Zrnic:PAWR_GARS,Mizutani:PAWR,Yoshikawa:CombBeam,Kitahara:PAWR_GRSS}, which is developed to detect hazardous weather phenomena.
The Doppler weather radar is not capable of detecting hazardous weather
because of its mechanical vertical scan for observing backscattered signals in multiple elevation angles,
which requires a long observation time.
To shorten the observation time,
the PAWR transmits a fan beam and subsequently
observes backscattered signals in a wide range of elevation angles.
The backscattered signals observed by the PAWR
can be modeled as mixtures of realizations of random processes
whose PSDs reflect the weather conditions in finely-divided ranges.
Thus, to obtain the weather condition in a fine resolution,
the PAWR needs digital signal processing, recovering the PSDs in the finely-divided ranges.

Since the estimation of PSDs from mixtures of realizations is a challenging problem,
the major existing methods employ a two-step approach
that first estimates the frequency components of the realizations
and then estimates the PSDs.
For the frequency component estimation,
sparsity-aware methods have achieved significant improvements on the estimation accuracy
over the classical linear methods in many fields \cite{Malioutov:SpartialSparse,Stoica:SpatialSparse,Wang:SpartialSparse,
Stoica:SparseLineSpectrum,Rojas:SparseLineSpectrum,Fang:SparseLineSpectrum,Shang:BayesSparse,
Kitahara:PAWR_GRSS}.
Sparsity in the spatial domain
is exploited in \cite{Malioutov:SpartialSparse,Stoica:SpatialSparse,Wang:SpartialSparse}
under the assumption that signals arrive from only a few angles.
Unfortunately, this assumption is far from suitable for the PAWR
because targets such as clouds and raindrops
exist at many angles \cite{Kitahara:PAWR_GRSS}.
In \cite{Stoica:SparseLineSpectrum,Rojas:SparseLineSpectrum,Fang:SparseLineSpectrum,Shang:BayesSparse},
isolated sparse frequency components, called \emph{line spectra},
are estimated based on the $\ell_1$ regularization.
It is demonstrated in \cite{Kitahara:PAWR_GRSS} that a block-sparse regularization model
using the mixed $\ell_2/\ell_1$ norm \cite{Yuan:l12,Stojnic:BlockSparse,Eldar:GroupSparse,Lv:GroupSparse,Elhamifar:GroupSparse,Jacob:l12,Obozinski:LGL:TechRep}
is more effective for weather radar applications
because the frequency components are clustered in a few blocks
due to the narrow-bandwidth of the PSDs.
After the frequency component estimation, the \emph{periodogram}, i.e.,
the squared magnitude of the frequency components,
is usually employed to estimate the PSD because of its asymptotic unbiasedness.
However, 
the periodogram has the drawbacks of
large variance and erratic oscillation \cite{Stoica:BookPSD,Brockwell:BookPSD,Oppenheim:BookDTSP,Bringi:BookDopplerRadar}.
For the classical case of a single random process,
smoothing techniques, e.g., those shown in \cite{Stoica:BookPSD,Brockwell:BookPSD}, are often used
to reduce the variance and the erratic oscillation.
While the existing smoothing techniques can be used as a post-processing step,
such a two-stage approach would be sub-optimal because
the smoothness is not exploited when estimating the frequency components.
Since the PSD is estimated by the periodogram, i.e., the squared magnitude of the frequency components,
one may add
a penalty for the difference between the magnitude of the frequency components in the frequency component estimation.
However, due to the nonconvexity of this type of penalty (see, e.g., \cite{Ciuciu:SmoothPSD}),
it is hard for this approach to obtain an optimal solution, and the performance dependency on the initial estimate and the optimization algorithm is difficult to elude.

Another line of studies derive approximated observation models
between the realizations and the PSD, e.g., for a
single random process \cite{Ariananda:SingleSource,Wang:SingleSource,Lu:ApproxCrossTerm,Zhao:ApproxCrossTerm}
and spatially independent random processes \cite{Bazerque:SpatialIndependent,Li:SpatialIndependent}.
Since the approximated observation model is written in terms of the (nonnegative) PSD,
the smoothness of the PSD could be exploited via convex optimization.
However, this approach takes the magnitude of the observation model to derive the approximated observation models,
implying that half of the information in the original observation model is lost as the phase information is discarded.
In particular, this approach is not applicable to the PAWR because signals from different angles cannot be distinguished by 
the magnitude information (see \eqref{eq:ObservationPAWR} in Example \ref{exmp:PAWR}).

In this paper, we propose a
convex optimization-based method that simultaneously estimates
block-sparse frequency components and
block-sparse and smooth PSDs from mixtures of realizations.\footnote{%
If a target is both sparse and smooth, it is also block-sparse since nonzero components are clustered in several blocks due to the smoothness.}
To design the proposed method,
we first apply the optimally structured block-sparse model of \cite{Kuroda:BlockSparse}
for the frequency component estimation. Then, we
newly leverage the latent variable of the designed model,
which is originally introduced to optimize the block structure, for the PSD estimation.
More precisely, we demonstrate that the latent variable is
in fact related to the square root of the PSDs,
enabling us to exploit the smoothness of the PSDs via convex optimization.
The main contributions of this paper are summarized as follows.
\begin{itemize}
\item We present, for the first time in the literature, a convex optimization-based method that can exploit the smoothness
of the PSDs for their estimation from mixtures of realizations.
\item We show that many smoothness priors designed for real-valued signals, including
the high-order total variation \cite{Chan:HTV,Maso:HTV} and the total generalized variation \cite{Bredies:TGB}, can be directly incorporated into the proposed framework
to enhance the smoothness of the PSDs of complex-valued random processes
thanks to the nonnegative latent variable.
\item We conduct thorough numerical simulations on the PAWR, which demonstrate that the proposed method achieves superior estimation accuracy to the existing sparse estimation models combined with or without post-smoothing, i.e., a smoothing technique applied as a post-processing step after the frequency component estimation.
\end{itemize}

The rest of this paper is organized as follows.
In Section \ref{sect:ProblemFormulation},
we formulate the estimation of PSDs from mixtures of realizations of random processes, and clarify its relation to weather radar applications.
In Section \ref{sect:ProposedApproach},
we design the proposed convex optimization model that
simultaneously estimates block-sparse frequency components
and block-sparse and smooth PSDs.
Section \ref{sect:experiment} presents numerical experiments on the PAWR, followed by
conclusion in Section \ref{sect:Conclusion}.

A preliminary short version of this paper was presented at a conference \cite{Kuroda:PSD_Conf}.

\noindent
{\it Notations}: $\mathbb{N}$, $\mathbb{R}$, $\mathbb{R}_{+}$, $\mathbb{R}_{++}$, and $\mathbb{C}$
respectively denote the sets of nonnegative integers, real numbers, nonnegative real numbers, positive real numbers, and complex numbers.
We use $\imath \in \mathbb{C}$ to denote the imaginary unit, i.e., $\imath = \sqrt{-1}$.
For every $x \in \mathbb{C}$,
$x^{*}$ denotes the complex conjugate of $x$,
and $|x| := \sqrt{x^{*} x}$ denotes the absolute value of $x$.
For matrices or vectors,
we denote the transpose and the Hermitian transpose respectively by $(\cdot)^\top$
and $(\cdot)^{\mathrm{H}}$.
The identity matrix of order $N$ is denoted by $\mathbf{I}_N \in \mathbb{R}^{N \times N}$.
We denote the diagonal matrix with components of $\mathbf{w} \in \mathbb{C}^{N}$ on the main diagonal
by $\mathrm{diag}(\mathbf{w}) \in \mathbb{C}^{N \times N}$.
The cardinality of a set $\mathcal{A}$ is denoted by $|\mathcal{A}|$.
The $\ell_2$ (Euclidean) norm, the $\ell_1$ norm,
and the $\ell_0$ pseudo-norm
of $\mathbf{x}  = (x_1,\ldots,x_N)^\top \in \mathbb{C}^{N}$ are respectively denoted by
$\|\mathbf{x}\| := \sqrt{\sum_{n=1}^{N}|x_n|^2}$,
$\|\mathbf{x}\|_1 := \sum_{n=1}^{N}|x_n|$,
and $\|\mathbf{x}\|_0 := \left|\{n \in \{1,\ldots,N\}\,|\, x_n \neq 0\}\right|$.
The expectation operator is denoted by
$E[\cdot]$.

\section{Problem Formulation}
\label{sect:ProblemFormulation}
We consider the estimation problem of power spectral densities (PSDs)
of $N$ random processes from noisy mixtures of their realizations.
We denote the $n$-th complex-valued discrete-time random process $(n = 1,\ldots,N)$ by
\begin{align}
\label{eq:def:random_process}
X_n^\star[\ell] \in \mathbb{C} \qquad (\ell = 0,\pm 1, \pm 2, \ldots).
\end{align}
To define the PSD of $X_n^\star[\ell]$, we assume that
$X_n^\star[\ell]$ is zero-mean and wide-sense stationary, which imply that
$E[X_n^\star[\ell]] = 0$ for any $\ell$, and
the auto-correlation $E[X_n^\star[m+\ell] (X_n^\star[m])^{*}]$
does not depend on $m$ for any $\ell$.
Under these assumptions, define the auto-correlation function by
\begin{align*}
R_n[\ell] := E[X_n^\star[m+\ell] (X_n^\star[m])^{*}],
\end{align*}
and suppose $\sum_{\ell=-\infty}^{\infty}|R_{n}[\ell]| < \infty$.
Then, the PSD of $X_n^\star[\ell]$ is given by
\begin{align}
\label{eq:def:DiscreteTimePSD}
S^{\star}_{n}(f) := \sum_{\ell=-\infty}^{\infty}R_{n}[\ell] e^{-\imath 2\pi f \ell} \quad 
\left(f \in \left[-\frac{1}{2},\frac{1}{2} \right)\right).
\end{align}
We denote $L$ consecutive realizations of $X_n^\star[\ell]$ by
\begin{align}
\label{eq:def:time_domain_realization}
\bar{x}_{j,n}[\ell] \in \mathbb{C} \qquad (\ell = 1,\ldots,L),
\end{align}
where $j \in \{1,\ldots,J\}$ is the index of trials.
Note that $\bar{x}_{j,n}[\ell]$ for $j=1,\ldots,J$
are assumed to be realizations of the common random process $X_n^\star[\ell]$
(see Remark \ref{rmrk:MultipleObservation} for validity of this assumption).
We define the observation model by
\begin{align}
\label{eq:def:ObservationModel_TimeDomain}
\mathbf{y}_{j} := \sum_{n=1}^{N}\mathbf{A}_n\bar{\mathbf{x}}_{j,n} + \boldsymbol{\varepsilon}_{j} \in \mathbb{C}^{d}
\quad (j = 1,\ldots,J),
\end{align}
where the realizations in \eqref{eq:def:time_domain_realization} are collectively denoted by
\begin{align}
\label{eq:def:time_domain_realization_vector}
\bar{\mathbf{x}}_{j,n} &:= ( \bar{x}_{j,n}[1],\bar{x}_{j,n}[2],\ldots,\bar{x}_{j,n}[L])^\top \in \mathbb{C}^{L},
\end{align}
$\mathbf{A}_n \in \mathbb{C}^{d\times L}$ is the known matrix that models the observation process for the $n$-th source,
and $\boldsymbol{\varepsilon}_{j} \in \mathbb{C}^d$ is the (unknown) observation noise.
Our goal is to estimate the PSDs $S^{\star}_{n}(f) \quad (n=1,\ldots,N)$
from $\mathbf{y}_{j} \quad (j=1,\ldots,J)$
and $\mathbf{A}_n\quad (n=1,\ldots,N)$ in \eqref{eq:def:ObservationModel_TimeDomain}.
Note that the classical PSD estimation problem for a single random process \cite{Stoica:BookPSD,Brockwell:BookPSD,Oppenheim:BookDTSP},
e.g., for the Doppler weather radar \cite{Bringi:BookDopplerRadar},
is a special instance of \eqref{eq:def:ObservationModel_TimeDomain} for $N = 1$,
$d=L$, and $\mathbf{A}_n = \mathbf{I}_{L}$.
The generalized observation model \eqref{eq:def:ObservationModel_TimeDomain} is introduced
to cover the PSD estimation for
the PAWR \cite{Zrnic:PAWR,McLaughlin:PAWR,Isom:PAWR,Yoshikawa:PAWR,Zrnic:PAWR_GARS,Mizutani:PAWR,Yoshikawa:CombBeam,Kitahara:PAWR_GRSS}, which is our primary interest.

\begin{exmp}[PAWR]
\label{exmp:PAWR}
For the PAWR, $X_n^\star[\ell]$ corresponds to the sum of backscattered signals in the angular interval
$\left[\theta_n-\frac{\Delta\theta}{2},\theta_n+\frac{\Delta\theta}{2}\right]$,
where $\theta_n \quad (n=1,\ldots,N)$ are the equally spaced angles with a spacing of $\Delta\theta$.
By using an $M$-element uniform linear array,
the PAWR observes noisy mixtures of realizations by
\begin{align}
\label{eq:ObservationPAWR}
\mathbf{y}_{j}[\ell] := \sum_{n=1}^{N}\mathbf{a}(\theta_{n})\bar{x}_{j,n}[\ell] + \boldsymbol{\varepsilon}_{j}[\ell] \in \mathbb{C}^{M} \quad (\ell=1,\ldots,L)
\end{align}
for each $j \in \{1,\ldots,J\}$, where
$\mathbf{a}(\theta_n)\in \mathbb{C}^{M}$ is the known steering vector for the angle $\theta_n$,
and $\boldsymbol{\varepsilon}_{j}[\ell] \in \mathbb{C}^{M}$ is the white Gaussian noise.
More precisely, $\mathbf{a}(\theta)$ is defined by
\begin{align*}
\mathbf{a}(\theta) := \left(1, e^{-\imath\frac{2\pi \Delta \sin \theta}{\lambda_{\mathrm{cw}}}},\ldots,e^{-\imath\frac{2(M-1)\pi \Delta \sin \theta}{\lambda_{\mathrm{cw}}}}\right)^\top \in \mathbb{C}^M,
\end{align*}
where $\Delta$ is the inter-element spacing of the uniform linear array,
and $\lambda_{\mathrm{cw}}$ is the carrier wavelength.
The observation model \eqref{eq:ObservationPAWR} for the PAWR can be written
in the form of \eqref{eq:def:ObservationModel_TimeDomain}, i.e.,
\begin{align*}
\mathbf{y}_{j}^{\mathrm{(pawr)}} = \sum_{n=1}^{N}\mathbf{A}_n^{\mathrm{(pawr)}}\bar{\mathbf{x}}_{j,n} + \boldsymbol{\varepsilon}_{j}^{\mathrm{(pawr)}}\quad (j = 1,\ldots,J),
\end{align*}
by setting
\begin{align*}
\mathbf{y}_j^{\mathrm{(pawr)}} &:= (\mathbf{y}_j[1]^\top,\mathbf{y}_j[2]^\top,\ldots,\mathbf{y}_j[L]^\top)^\top \in \mathbb{C}^{ML},\\
\boldsymbol{\varepsilon}_j^{\mathrm{(pawr)}} &:= (\boldsymbol{\varepsilon}_j[1]^\top,\boldsymbol{\varepsilon}_j[2]^\top,\ldots,\boldsymbol{\varepsilon}_j[L]^\top)^\top \in \mathbb{C}^{ML},
\end{align*}
and $\mathbf{A}_n^{\mathrm{(pawr)}} \in \mathbb{C}^{ML \times L}$
to the block-diagonal matrix that contains $L$ copies of $\mathbf{a}(\theta_n)$
on the diagonal blocks, i.e.,
\begin{align*}
\mathbf{A}_n^{\mathrm{(pawr)}} :=& \begin{pmatrix}
\mathbf{a}(\theta_n) & & &\\
& \mathbf{a}(\theta_n) & &\\
& &\ddots &\\
& & &\mathbf{a}(\theta_n)
\end{pmatrix}.
\end{align*}
The estimation of the PSDs $S^{\star}_{n}(f)$ of $X_n^\star[\ell] \quad (n=1,\ldots,N)$ is essential
for the PAWR because $S^{\star}_{n}(f)$ contains information about the weather condition in the narrow angular interval $\left[\theta_n-\frac{\Delta\theta}{2},\theta_n+\frac{\Delta\theta}{2}\right]$.
More precisely, the weather condition can be obtained from $S^{\star}_{n}(f)$ as follows.
First, the continuous-time PSD $S_n^{\star \mathrm{(ct)}}(f)$ is recovered from
$S^{\star}_{n}(f)$.
When aliasing does not occur in $S^{\star}_{n}(f)$, $S_n^{\star \mathrm{(ct)}}(f)$ can be simply obtained
by $S_n^{\star \mathrm{(ct)}}(f) = T S_n^{\star}(T f)$ if $|f| \leq \frac{1}{2T}$
and $S_n^{\star \mathrm{(ct)}}(f) = 0$ otherwise, where $T$ is the pulse repetition time, i.e.,
the sampling interval of $X_n^\star[\ell]$.
Even when aliasing occurs in $S^{\star}_{n}(f)$,
$S_n^{\star \mathrm{(ct)}}(f)$ can be recovered from
$S^{\star}_{n}(f)$ unless the variation of wind velocity is extremely large,
and thus the anti-aliasing filter is usually not employed in weather radar applications
(see, e.g., \cite[Chapter 5]{Bringi:BookDopplerRadar} and \cite[Section II-A]{Kitahara:PAWR_GRSS} for detail). Next, $S_n^{\star \mathrm{(ct)}}(f)$ is decomposed as
\begin{align*}
S_n^{\star \mathrm{(ct)}}(f) = P_n^{\star}q_n^{\star}(f),
\end{align*}
where $P_n^{\star} := \int_{-\infty}^{\infty}S_n^{\star \mathrm{(ct)}}(f) df$
and $q_n^{\star}(f) := S_n^{\star \mathrm{(ct)}}(f)/P_n^{\star}$
respectively correspond to the precipitation intensity and the Doppler frequency distribution \cite{Bringi:BookDopplerRadar}.
The Doppler frequency distribution
can be converted to the distribution of Doppler velocity $v$, i.e., wind velocity parallel to the incident beam direction, by $v = \frac{\lambda_{\mathrm{cw}} f}{2}$. For instance, the area of nonzero values of $q_n^{\star}(f)$
implies the existence of the corresponding wind velocity components.
The precipitation intensity and the Doppler velocity distribution
are useful for the analysis of weather phenomena, e.g.,
\cite{Zrnic:AnalysisPSD,Wang:AnalsisPSD,Yu:AnalysisPSD,Bluestein:AnalysisPSD} for tornado,
\cite{Janssen:AnalysisPSD} for weather clutter, and
\cite{Yoshikawa:AnalysisPSD} for raindrop size distribution.
Since the estimation accuracy of the precipitation intensity and the Doppler velocity distribution
heavily depends on that of the PSD,
it is important for the analysis of weather phenomena based on the PAWR to realize a method that can
accurately estimate the PSDs $S^{\star}_{n}(f) \quad (n=1,\ldots,N)$ from the mixtures of realizations.
\end{exmp}

\begin{rmrk}[Tradeoff between $L$ and $J$]
\label{rmrk:MultipleObservation}
To derive the observation model in \eqref{eq:def:ObservationModel_TimeDomain}
where $\bar{x}_{j,n}[\ell]\quad(j=1,\ldots,J)$ are realizations of the common random process $X_n^\star[\ell]$,
similarly to the case of a single random process $(N=1)$
\cite{Stoica:BookPSD,Brockwell:BookPSD,Oppenheim:BookDTSP,Bringi:BookDopplerRadar},
we split all observations into $J$ subsets.
For weather radar applications,
the total number $K_{\mathrm{pls}}$ of pulses is divided
into $J$ subsets, and thus we have $L = \frac{K_{\mathrm{pls}}}{J}$.
Since $\frac{1}{L} = \frac{J}{K_{\mathrm{pls}}}$ is the frequency resolution, i.e.,
the sampling interval in the frequency domain (see \eqref{eq:def:frequency_grid}),
increasing $J$ sacrifices the frequency resolution \cite[Chapter 5]{Bringi:BookDopplerRadar}.
Note that we cannot increase $K_{\mathrm{pls}}$ unboundedly
because $K_{\mathrm{pls}}$ corresponds to the observation time,
and thus is set to be small enough
to ensure that the statistics of targets such as clouds and raindrops are (approximately) unchanged.
Typically, $J$ is set to be very small for the sake of a fine frequency resolution, and
$J = 1$ is of particular interest as the original frequency resolution $\frac{1}{K_{\mathrm{pls}}}$ is preserved \cite{Bringi:BookDopplerRadar}.
Note that, while $J=1$ is a typical choice in practice,
our method, which will be developed in Section \ref{sect:ProposedApproach}, is applicable to general $J$.
\end{rmrk}

We rewrite \eqref{eq:def:ObservationModel_TimeDomain} to an observation model in terms of
frequency components of the time-domain realizations $\bar{\mathbf{x}}_{j,n}$ in \eqref{eq:def:time_domain_realization_vector}
because of their more direct relation to the PSD in \eqref{eq:def:DiscreteTimePSD} than
$\bar{\mathbf{x}}_{j,n}$.
More precisely, we represent $\bar{\mathbf{x}}_{j,n}$ as
\begin{align}
\label{eq:def:frequency_representation_with_dictionary}
\bar{\mathbf{x}}_{j,n} = \mathbf{G}\bar{\mathbf{u}}_{j,n}
\end{align}
for $j = 1,\ldots,J$ and $n = 1,\ldots,N$, where
\begin{align}
\label{eq:def:frequency_components}
\bar{\mathbf{u}}_{j,n} :=(\bar{u}_{j,n}[1],\bar{u}_{j,n}[2],\ldots,\bar{u}_{j,n}[L])^\top \in \mathbb{C}^{L}
\end{align}
is used as the frequency components,
and $\mathbf{G} \in \mathbb{C}^{L \times L}$ is a suitable synthesis matrix.
Substituting the representation \eqref{eq:def:frequency_representation_with_dictionary} to
the observation model \eqref{eq:def:ObservationModel_TimeDomain}, we have
\begin{align}
\label{eq:GeneralObservationModelFreqDom}
\mathbf{y}_{j} = \sum_{n=1}^{N}\mathbf{A}_n\mathbf{G}\bar{\mathbf{u}}_{j,n} + \boldsymbol{\varepsilon}_{j} \in \mathbb{C}^{d}
\quad (j = 1,\ldots,J),
\end{align}
which is used as the observation model for the frequency components $\bar{\mathbf{u}}_{j,n}$.
The representation \eqref{eq:def:frequency_representation_with_dictionary}
covers popular frequency analysis methods used in weather radar applications, e.g.,
the discrete Fourier transform (DFT) and the windowed DFT.

\begin{exmp}[DFT]
\label{exmp:DFT}
Let $\bar{\mathbf{u}}^{\mathrm{(DFT)}}_{j,n}$ be the normalized DFT coefficients of $\bar{\mathbf{x}}_{j,n}$.
Define the normalized DFT matrix $\mathbf{F} \in \mathbb{C}^{L \times L}$ by
\begin{align}
\label{eq:def:DFTmatrix}
\mathbf{F} := \frac{1}{\sqrt{L}}\begin{pmatrix}
1 &e^{-\imath 2\pi f_1} & \cdots &e^{-\imath 2\pi f_1 (L-1)}\\
1 &e^{-\imath 2\pi f_2} & \cdots &e^{-\imath 2\pi f_2 (L-1)}\\
\vdots&\vdots &\ddots &\vdots\\
1 &e^{-\imath 2\pi f_L} & \cdots &e^{-\imath 2\pi f_L (L-1)}
\end{pmatrix},
\end{align}
where
\begin{align}
\label{eq:def:frequency_grid}
f_k := \frac{k-1-L/2}{L} \quad (k=1,\ldots,L)
\end{align}
are uniform sampling points in $[-1/2,1/2)$ used as a frequency grid, and $L$ is assumed to be even for simplicity.
Then, $\bar{\mathbf{u}}_{j,n}^{\mathrm{(DFT)}}$ is given by
\begin{align*}
\bar{\mathbf{u}}_{j,n}^{\mathrm{(DFT)}} = \mathbf{F}\bar{\mathbf{x}}_{j,n}.
\end{align*}
Due to the unitarity of $\mathbf{F}$, we have
\begin{align*}
\bar{\mathbf{x}}_{j,n} = \mathbf{F}^{\mathrm{H}}\bar{\mathbf{u}}_{j,n}^{\mathrm{(DFT)}},
\end{align*}
which corresponds to \eqref{eq:def:frequency_representation_with_dictionary} with $\mathbf{G} = \mathbf{F}^{\mathrm{H}}$.
\end{exmp}

\begin{exmp}[Windowed DFT]
\label{exmp:windowedDFT}
To mitigate the frequency sidelobes, a window function
$\mathbf{w} \in \mathbb{R}_{++}^{L}$ is applied to $\bar{\mathbf{x}}_{j,n}$ before the DFT
in some cases \cite{Oppenheim:BookDTSP,Chandrasekar:BookDopplerRadar}. The windowed DFT coefficients are given by
\begin{align*}
\bar{\mathbf{u}}_{j,n}^{\mathrm{(WDFT)}} = \mathbf{F}\mathbf{W}\bar{\mathbf{x}}_{j,n},
\end{align*}
where $\mathbf{F}$ is the DFT matrix in \eqref{eq:def:DFTmatrix}, and
$\mathbf{W} := \mathrm{diag}(\mathbf{w}) \in \mathbb{R}_{++}^{L \times L}$.
Since $(\mathbf{F}\mathbf{W})^{-1} = \mathbf{W}^{-1}\mathbf{F}^{\mathrm{H}}$,
we have
\begin{align*}
\bar{\mathbf{x}}_{j,n} = \mathbf{W}^{-1}\mathbf{F}^{\mathrm{H}}\bar{\mathbf{u}}_{j,n}^{\mathrm{(WDFT)}},
\end{align*}
which corresponds to \eqref{eq:def:frequency_representation_with_dictionary} with
$\mathbf{G} = \mathbf{W}^{-1}\mathbf{F}^{\mathrm{H}}$.
\end{exmp}

\subsection*{Major Challenge in PSD Estimation from Mixtures of Realizations}
The square of the magnitude of the frequency components in \eqref{eq:def:frequency_components}, i.e.,
\begin{align}
\label{eq:def:periodogram}
|\bar{u}_{j,n}[k]|^2 \quad (k=1,\ldots,L),
\end{align}
is called the \emph{periodogram}\footnote{%
Strictly speaking, \eqref{eq:def:periodogram} is called the periodogram when
$\mathbf{G}$ in \eqref{eq:def:frequency_representation_with_dictionary} is the inverse DFT $\mathbf{F}^{\mathrm{H}}$
in Example \ref{exmp:DFT}, and \eqref{eq:def:periodogram} with $\mathbf{G} = \mathbf{W}^{-1}\mathbf{F}^{\mathrm{H}}$
is called, e.g., the \emph{windowed periodogram} \cite{Stoica:BookPSD} or the \emph{modified periodogram} \cite{Oppenheim:BookDTSP}.} and
widely used as an estimate of the PSD $S^{\star}_{n}(f)$ on the frequency grid defined in \eqref{eq:def:frequency_grid}.
It should be noted that
the periodogram needs to be estimated from the mixtures of realizations in \eqref{eq:GeneralObservationModelFreqDom} for our problem.
The periodogram with the DFT shown in Example \ref{exmp:DFT}
is an asymptotically unbiased estimator of the PSD
under a mild condition \cite{Stoica:BookPSD,Brockwell:BookPSD}.
More precisely, since $\bar{u}_{j,n}^{\mathrm{(DFT)}}[k]$ can be regarded as a realization of the random variable
\begin{align*}
U_{n}^{\star}[k] := \frac{1}{\sqrt{L}}\sum_{\ell=1}^{L}X_n^\star[\ell] e^{-\imath 2\pi f_k (\ell-1)}\quad (k=1,\ldots,L),
\end{align*}
the asymptotic unbiasedness means that
\begin{align}
\label{eq:ExpectationPeriodogram}
\lim_{L \rightarrow \infty} E\left[|U_{n}^{\star}[k]|^2\right] = S^{\star}_{n}(f_k) \quad (k=1,\ldots,L).
\end{align}
A simple proof of \eqref{eq:ExpectationPeriodogram}
is provided in \cite[Section II-A]{Kitahara:PAWR_GRSS} under a mild sufficient condition $\sum_{\ell=-\infty}^{\infty}|\ell R_{n}[\ell]| < \infty$.
The periodogram with the windowed DFT in Example \ref{exmp:windowedDFT}
is also asymptotically unbiased if the window function is properly designed \cite{Oppenheim:BookDTSP}.

\begin{figure}[t]
  \centering
    \includegraphics[width=\columnwidth]{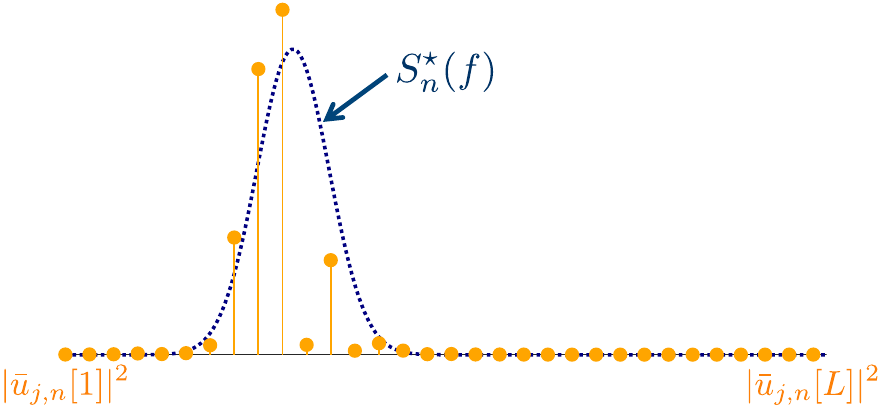}
  \caption{The PSD and the periodogram computed from the realizations for a simulation.}
  \label{fig:illustPSDandPeriodogram}
\end{figure}

The drawback of the periodogram lies in
its large variance and often-observed erratic oscillation \cite{Stoica:BookPSD,Brockwell:BookPSD,Oppenheim:BookDTSP,Bringi:BookDopplerRadar}
(see also Fig.~\ref{fig:illustPSDandPeriodogram} for an example from the experiments in Section \ref{sect:experiment}).
Indeed, this drawback is theoretically shown
for the periodogram with the DFT
when $X_n^\star[\ell]$ is a linear combination of i.i.d.~Gaussian random variables and $L \rightarrow \infty$ \cite{Stoica:BookPSD,Brockwell:BookPSD}.
Namely, for this case,
the variance is equally large to the square of the PSD, i.e.,
\begin{align*}
\lim_{L \rightarrow \infty} E\left[\left(|U_{n}^{\star}[k]|^2 - S^{\star}_{n}(f_k) \right)^2\right] = (S^{\star}_{n}(f_k))^2
\end{align*}
for $k=1,\ldots,L$, and $U_{n}^{\star}[k]$ and $U_{n}^{\star}[k']$ $(k \neq k')$ are uncorrelated
when $L \rightarrow \infty$. These facts validate the often-observed erratic oscillation of the periodogram.

Although several approaches have been developed to reduce the variance of the periodogram,
they are not suitable for our problem in which the PSDs have to be estimated from the mixtures of realizations in \eqref{eq:GeneralObservationModelFreqDom}.
A simple approach is to
exploit the situation that $\bar{u}_{j,n}[k] \quad (j=1,\ldots,J)$ are realizations of the common random variable,
i.e., to use the ensemble average of the periodograms:
\begin{align}
\label{eq:AveragedPeriodogram}
\frac{1}{J}\sum_{j=1}^{J}|\bar{u}_{j,n}[k]|^2 \quad (k=1,\ldots,L).
\end{align}
Unfortunately, since $J$ is typically very small in weather radar applications (see Remark \ref{rmrk:MultipleObservation}),
the ensemble average cannot sufficiently reduce the variance and the erratic oscillation.
Since the PSD is usually smooth in weather radar applications \cite{Bringi:BookDopplerRadar},
another promising approach is to exploit
the smoothness of the PSD.
However, existing smoothing techniques, e.g., those shown in
\cite{Stoica:BookPSD,Brockwell:BookPSD},
are not directly applicable to our problem because 
these techniques suppose that the frequency components $\bar{\mathbf{u}}_{j,n}$ of the realizations are
known.
Using smoothing techniques
as a post-processing step
would be sub-optimal because
the smoothness of the PSD is not considered in the estimation of the frequency components.
Thus, it remains a major challenge to exploit the smoothness
of the PSDs when they need to be estimated from mixtures of realizations.

\section{Proposed Approach}
\label{sect:ProposedApproach}
To exploit the sparsity and the smoothness for the PSD estimation
from the observed mixtures of realizations in \eqref{eq:GeneralObservationModelFreqDom},
we design a convex model that jointly estimates the frequency components $\bar{u}_{j,n}[k]$ and
the PSDs $S^{\star}_{n}(f_k)$.
In Section \ref{sect:PropFreqEst}, we first apply the optimally structured block-sparse model of \cite{Kuroda:BlockSparse}
for the estimation of the frequency components.
Then, in Section \ref{sect:PropPSDEst}, we leverage its latent variable,
which is originally introduced for the block structure optimization,
to estimate sparse and smooth PSDs.

\subsection{Block-Sparse Estimation of Frequency Components}
\label{sect:PropFreqEst}
\begin{figure}[t]
  \centering
    \includegraphics[width=\columnwidth]{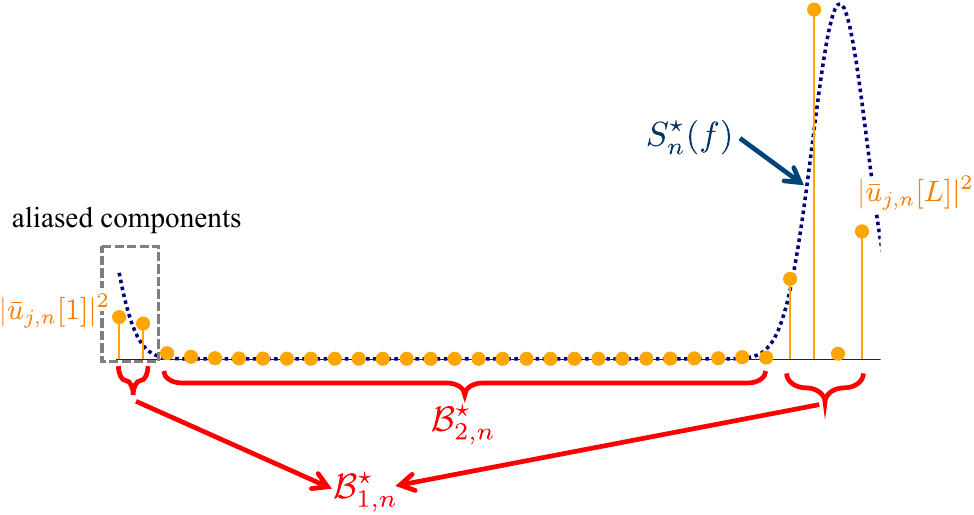}
  \caption{The PSD and the squared magnitude of the frequency components when aliasing occurs, and its suitable block partition.}
  \label{fig:illustPSDandPeriodogram_Aliasing}
\end{figure}

We design a block-sparse penalty for the
frequency components $\bar{\mathbf{u}}_{j,n}$
by applying the optimally structured block-sparse model \cite{Kuroda:BlockSparse}
with the knowledge of the PAWR \cite{Kitahara:PAWR_GRSS}.
For simplicity, we begin by designing a penalty for each $n \in \{1,\ldots,N\}$.
As demonstrated in \cite{Kitahara:PAWR_GRSS},
the PSD $S^{\star}_{n}(f)$ is usually narrow-band
for the PAWR, which implies that
$\bar{\mathbf{u}}_{j,n}$ is block-sparse for each source $n\in\{1,\ldots,N\}$ and trial $j\in\{1,\ldots,J\}$
due to the relation \eqref{eq:ExpectationPeriodogram}.\footnote{%
Although we present \eqref{eq:ExpectationPeriodogram} for $\mathbf{G} = \mathbf{F}^{\mathrm{H}}$ in Example \ref{exmp:DFT} for simplicity,
$\bar{\mathbf{u}}_{j,n}$ is also block-sparse when $\mathbf{G} = \mathbf{W}^{-1}\mathbf{F}^{\mathrm{H}}$ in Example \ref{exmp:windowedDFT}
because the window function in Example \ref{exmp:windowedDFT} is designed to
reduce the heights of the sidelobes
and slightly increase the width of the mainlobe.}
Moreover, since $\bar{\mathbf{u}}_{j,n} \quad (j=1,\ldots,J)$ are realizations of
the common random variable, suitable block partitions
for $\bar{\mathbf{u}}_{j,n} \quad (j=1,\ldots,J)$ are the same.
Thus, using the mixed $\ell_2/\ell_1$ norm that is suitable for the block-sparsity,
we introduce
a penalty for $\mathbf{u}_{n} := (\mathbf{u}_{j,n})_{j=1}^J$ as
\begin{align*}
\left\|\mathbf{u}_{n} \right\|_{2,1}^{(\mathcal{B}_{m,n})_{m=1}^{h_{n}}}
:=& \sum_{m=1}^{h_{n}}\sqrt{J|\mathcal{B}_{m,n}|}
\left\| ((u_{j,n}[k])_{j=1}^J)_{k \in \mathcal{B}_{m,n}}\right\|\\
=& \sum_{m=1}^{h_{n}}\sqrt{J|\mathcal{B}_{m,n}|}
\sqrt{\sum_{j=1}^{J}\sum_{k \in \mathcal{B}_{m,n}} |u_{j,n}[k]|^2 },
\end{align*}
where $\mathcal{B}_{m,n} \subset \{1,\ldots,L\}\quad(m=1,\ldots,h_{n})$ is a block partition in the frequency domain of the $n$-th source.
By suppressing the mixed $\ell_{2}/\ell_{1}$ norm, the block-sparsity is promoted because the components
$((u_{j,n}[k])_{j=1}^J)_{k \in \mathcal{B}_{m,n}}$ in the same block are forced to be zeros together.
The problem in \cite{Kitahara:PAWR_GRSS} is that an appropriate block partition is unknown a priori
because it depends on the unknown Doppler velocity distribution.
To solve the problem of unknown block partition, following the approach of \cite{Kuroda:BlockSparse}, we minimize
the mixed $\ell_2/\ell_1$ norm over the partition of at most $H_{n}$ blocks, i.e.,
\begin{align}
\label{eq:def:nonconvexPenalty}
\psi_{H_{n}}(\mathbf{u}_{n})
:=\min_{h_{n} \in \{1,\ldots,H_{n}\}}
\left[\min_{\left(\mathcal{B}_{m,n}\right)_{m=1}^{h_{n}} \in \mathcal{P}_{h_{n}}} \|\mathbf{u}_n \|_{2,1}^{(\mathcal{B}_{m,n})_{m=1}^{h_{n}}}\right].
\end{align}
The constraint set $\mathcal{P}_{h_{n}}$ consists of all $h_{n}$ block partitions of $\{1,\ldots,L\}$, i.e.,
\begin{equation*}
\begin{aligned}
&\left(\mathcal{B}_{m,n}\right)_{m=1}^{h_{n}} \in \mathcal{P}_{h_{n}} \\ \Leftrightarrow
&\left\{\begin{aligned}
&\bigcup_{m=1}^{h_{n}}\mathcal{B}_{m,n} = \{1,\ldots,L\},\\
&\mathcal{B}_{m,n} \neq \varnothing  \quad  (m = 1,\ldots,h_{n}),\\
&\mathcal{B}_{m,n} \cap \mathcal{B}_{m',n} = \varnothing  \quad  (m \neq m'),\\
&\mathcal{B}_{m,n} = \mathfrak{M}_{L}\left(\{\ell \in \mathbb{N} \mid a_{m,n} \leq \ell \leq b_{m,n} \}\right)
\\&\mbox{ for some } a_{m,n}, b_{m,n} \in \mathbb{N} \quad  (m = 1,\ldots,h_{n}),
\end{aligned}\right.\end{aligned}
\end{equation*}
where 
\begin{align*}
\mathfrak{M}_{L}(\mathcal{I}) := \left\{\left.\ell - L\left\lfloor\frac{\ell-1}{L}\right\rfloor  \in \{1,\ldots,L \} \,\right|\, \ell \in \mathcal{I}\right\},
\end{align*}
and $\lfloor\cdot \rfloor$ is the floor function.
For instance, when $\mathcal{I} = \{L-1,L,L+1\}$,
$\mathfrak{M}_{L}(\mathcal{I}) = \{L-1,L,1\}$.
Differently from the standard design proposed in \cite{Kuroda:BlockSparse},
the present design makes $\mathcal{P}_{h_{n}}$ includes blocks connected by the first and the last entries,
and is suitable for weather radar applications due to the following reason.
Since aliasing is not a serious issue in weather radar applications, the anti-aliasing filter is usually not employed (see Example \ref{exmp:PAWR} and references \cite[Chapter 5]{Bringi:BookDopplerRadar} and \cite[Section II-A]{Kitahara:PAWR_GRSS} for detail),
and thus aliasing may occur, i.e., some Doppler frequency components may exceed the Nyquist frequency.
For instance, in Fig.~\ref{fig:illustPSDandPeriodogram_Aliasing},
a part of the PSD that exceeds the Nyquist frequency is aliased, and thus the corresponding frequency components are also aliased.
In such cases, aliased nonzero components and non-aliased nonzero components are better to be collected into a single block, as shown in Fig.~\ref{fig:illustPSDandPeriodogram_Aliasing}, because they form a single block before the aliasing.
To realize such capability, $\mathcal{P}_{h_{n}}$ is designed to include blocks connected by the first and the last entries.
Note that a block connected by the first and the last entries is not always adopted
since the block partition is automatically optimized in \eqref{eq:def:nonconvexPenalty}.

Although it is difficult to use $\psi_{H_{n}}(\mathbf{u}_{n})$ directly due to the combinatorial optimization in \eqref{eq:def:nonconvexPenalty},
we can construct a tight convex relaxation of $\psi_{H_{n}}(\mathbf{u}_{n})$ as follows.
Let $\phi\colon\mathbb{C}^J\times\mathbb{R}_{+}\rightarrow\mathbb{R}_{+}\cup{\{\infty\}}$ be 
a lower semicontinuous convex function defined by
\begin{align}
\label{eq:defenition:phi}
\phi(\mathbf{v},\sigma) := \begin{dcases}
\frac{\| \mathbf{v}\|^2}{2\sigma} + \frac{J}{2}\sigma, &\mbox{{\rm if} }\sigma > 0;\\
0, &\mbox{{\rm if} } \mathbf{v}=\boldsymbol{0} \mbox{ {\rm and} } \sigma = 0;\\
\infty, &\mbox{{\rm otherwise}}.
\end{dcases}
\end{align}
Then, similarly to \cite[Section II]{Kuroda:BlockSparse},
$\psi_{H_{n}}(\mathbf{u}_{n})$ can be rewritten as
\begin{align}
\label{eq:nonconvexPenaltyL0Cons}
\psi_{H_{n}}\left(\mathbf{u}_{n}\right) = \min_{\substack{\boldsymbol{\sigma}_n \in \mathbb{R}_{+}^L \vspace{1.5pt} \\ \|\mathbf{D}\boldsymbol{\sigma}_n \|_{0} \leq H_{n}}} \sum_{k = 1}^{L} \phi \left( (u_{j,n}[k])_{j=1}^{J},\sigma_n[k]\right),
\end{align}
where $\mathbf{D} \in \mathbb{R}^{L \times L}$ is the first-order difference operator
with the periodic boundary condition, i.e., the difference operator on the
ring graph \cite{Kuroda:GraphSpares}. More precisely, $\mathbf{D}$ is defined by
\begin{align}
\label{eq:def:FirstDifferenceOp}
\mathbf{D} := \begin{bmatrix}
-1&  1 & 0& 0 & \cdots & 0& 0\\
0 & -1 & 1& 0& \cdots & 0& 0 \\
\vdots &\vdots & \vdots &  \vdots & \ddots &\vdots & \vdots \\
0 & 0 & 0 & 0 & \cdots & -1 & 1\\
1 & 0 & 0 & 0 & \cdots & 0 & -1
\end{bmatrix} \in \mathbb{R}^{L \times L}.
\end{align}
Note that \eqref{eq:nonconvexPenaltyL0Cons} is a slight extension of the result shown in \cite{Kuroda:BlockSparse} to the case where the blocks are fixed over the trials $j\in\{1,\ldots,J\}$.
We can obtain a tight convex relaxation of \eqref{eq:nonconvexPenaltyL0Cons}
by replacing the $\ell_0$ pseudo-norm in the constraint with its best convex relaxation, i.e., the $\ell_1$ norm:
\begin{align*}
\tilde{\psi}_{\alpha_n}(\mathbf{u}_{n}) := \min_{\substack{\boldsymbol{\sigma}_n \in \mathbb{R}_{+}^L \vspace{1.5pt} \\ \|\mathbf{D}\boldsymbol{\sigma}_n \|_{1} \leq \alpha_n}} \sum_{k = 1}^{L} \phi \left( (u_{j,n}[k])_{j=1}^{J},\sigma_n[k]\right),
\end{align*}
where $\alpha_n \in \mathbb{R}_{+}$ is a tuning parameter related to the number of blocks.
Although the sum $\sum_{n=1}^{N}\tilde{\psi}_{\alpha_n}(\mathbf{u}_{n})$ can be used
for the penalty of $(\mathbf{u}_{n})_{n=1}^{N}$,
tuning $\alpha_n$ for each $n \in \{1,\ldots,N\}$ could be troublesome.
Thus,
to simplify the tuning process, we propose a convex penalty for $\mathbf{u} := (\mathbf{u}_{n})_{n=1}^{N}$ as
\begin{align}
\label{eq:def:ProposedConvexPenalty}
\Psi_{\alpha}(\mathbf{u})
 :=\min_{\substack{(\boldsymbol{\sigma}_{n})_{n=1}^{N} \in \mathbb{R}_{+}^{NL} \vspace{1.5pt} \\ 
\sum_{n=1}^{N}\|\mathbf{D}\boldsymbol{\sigma}_n \|_{1} \leq \alpha}}
\sum_{n=1}^{N}\sum_{k = 1}^{L} \phi \left( (u_{j,n}[k])_{j=1}^{J},\sigma_n[k]\right),
\end{align}
where the single tuning parameter $\alpha \in \mathbb{R}_{+}$ is related to the number of total blocks
for $n = 1,\ldots,N$.

Using the proposed convex penalty $\Psi_{\alpha}(\mathbf{u})$ in \eqref{eq:def:ProposedConvexPenalty},
we estimate the frequency components
by the regularized least squares for the observation model in \eqref{eq:GeneralObservationModelFreqDom}, i.e.,
\begin{align}
\label{eq:PropRegDirectForm}
\minimize_{\mathbf{u} \in \mathbb{C}^{JNL}}
\frac{1}{2}\sum_{j=1}^{J}
\left\|\mathbf{y}_j-\sum_{n=1}^{N}\mathbf{A}_n\mathbf{G}\mathbf{u}_{j,n}\right\|^2 + \lambda\Psi_{\alpha}\left(\mathbf{u}\right),
\end{align}
where $\lambda > 0$ is the regularization parameter that controls the importance of the block-sparsity.
Substituting the definition of $\Psi_{\alpha}\left(\mathbf{u}\right)$  in \eqref{eq:def:ProposedConvexPenalty} into \eqref{eq:PropRegDirectForm},
we can reduce the optimization problem \eqref{eq:PropRegDirectForm} to
\begin{equation}
\label{eq:PropReg}
\left.\begin{aligned}
\minimize_{\mathbf{u} \in \mathbb{C}^{JNL}, \boldsymbol{\sigma} \in \mathbb{R}_{+}^{NL}}
\frac{1}{2}&\sum_{j=1}^{J}
\left\|\mathbf{y}_j-\sum_{n=1}^{N}\mathbf{A}_n\mathbf{G}\mathbf{u}_{j,n}\right\|^2\\
+&\lambda\sum_{n=1}^{N}\sum_{k = 1}^{L} \phi \left( (u_{j,n}[k])_{j=1}^{J},\sigma_n[k]\right)&\\
&\mbox{subject to }  \sum_{n=1}^{N}\|\mathbf{D}\boldsymbol{\sigma}_n \|_{1}  \leq \alpha
\end{aligned}\right\},
\end{equation}
where $\boldsymbol{\sigma}$ denotes $(\boldsymbol{\sigma}_{n})_{n=1}^{N}$.
Although the proposed regularization model \eqref{eq:PropReg}
is a relatively difficult convex optimization problem due to
the discontinuous function $\phi$,
we can obtain a globally optimal solution of \eqref{eq:PropReg}
by applying the
proximal splitting techniques \cite{Chambolle:PDS,Gabay:ADMM,Eckstein:ADMM,Combettes:ProxSplit,Condat:prox}
with the interpretation of $\phi$ in \eqref{eq:defenition:phi} as a \emph{perspective function}\cite{Combettes:Perspective,Combettes:PerspectiveProx,Combettes:PerspectiveML}.
A concrete algorithm based on the alternating direction method of multipliers (ADMM)
and its derivation are provided in Appendix \ref{appendix:OptimizationAlgorithm}.

\subsection{Leveraging Latent Variable for PSD Estimation}
\label{sect:PropPSDEst}
We demonstrate that the solution for the latent variable $\boldsymbol{\sigma}$
of the proposed model \eqref{eq:PropReg} is in fact suitable for the PSD estimation.
Let $\hat{\mathbf{u}}$ and $\hat{\boldsymbol{\sigma}}$ be
the solutions of \eqref{eq:PropReg}
respectively for the variables $\mathbf{u}$ and $\boldsymbol{\sigma}$.
While it is possible to compute the periodogram as in \eqref{eq:AveragedPeriodogram} 
for $\hat{\mathbf{u}}$, 
$\hat{\boldsymbol{\sigma}}$ is more suitable for the estimation of smooth PSDs.
To confirm this, we show that
$\hat{\boldsymbol{\sigma}}$ corresponds to the square root of smoothed and averaged periodogram as follows.
\begin{enumerate}
\item[i)] We begin by considering the case of $\alpha \rightarrow \infty$, which is not of our interest but easy to analyze.
In this case,
since $\boldsymbol{\sigma}$ minimizes
\begin{align}
\label{eq:PenaltyPhiPart}
\lambda\sum_{n=1}^{N}\sum_{k = 1}^{L} \phi \left( (u_{j,n}[k])_{j=1}^{J},\sigma_n[k]\right)
\end{align}
in \eqref{eq:PropReg},
the solutions $\hat{\mathbf{u}}$ and $\hat{\boldsymbol{\sigma}}$ satisfy the relationship
\begin{align}
\label{eq:RelationSigmaToAvePerio}
\hat{\sigma}_n[k] = \sqrt{\frac{1}{J}\sum_{j=1}^{J}|\hat{u}_{j,n}[k]|^2}
\end{align}
for each $k=1,\ldots,L$ and $n=1,\ldots,N$,
which can be shown based on \cite[Lemma 1]{Kuroda:BlockSparse}.
The relation \eqref{eq:RelationSigmaToAvePerio} means that
$\hat{\boldsymbol{\sigma}}$ is the square root of the averaged periodogram
in \eqref{eq:AveragedPeriodogram} computed with $\hat{\mathbf{u}}$ when $\alpha \rightarrow \infty$.
\item[ii)] 
Next, we consider the case of our interest, where
$\alpha$ is set to a finite value.
In this case, the constraint
\begin{align*}
\sum_{n=1}^{N}\|\mathbf{D}\boldsymbol{\sigma}_n \|_{1}  \leq \alpha
\end{align*}
in \eqref{eq:PropReg} penalizes the smoothness of
$\boldsymbol{\sigma}_{n} \quad (n=1\,\ldots,N)$ since $\mathbf{D}$ is the difference operator.
Meanwhile,
the other part \eqref{eq:PenaltyPhiPart} of the proposed model forces
$\boldsymbol{\sigma}$ to be the averaged periodogram 
in \eqref{eq:RelationSigmaToAvePerio}.
Thus, by the combination of these terms, roughly speaking,
$\hat{\boldsymbol{\sigma}}$ is 
smoothed around
the square root of the averaged periodogram in \eqref{eq:RelationSigmaToAvePerio}.
\end{enumerate}

In addition to being smooth, $\hat{\boldsymbol{\sigma}}$ is block-sparse
because $\hat{\mathbf{u}}_{j,n}$ $(j=1,\ldots,J)$ are regularized to have a common block-sparse support.
Since the PSDs are smooth and block-sparse for the PAWR,
the square of components of $\hat{\boldsymbol{\sigma}}$, i.e.,
\begin{align}
\label{eq:ProposedEstimatePSD}
\hat{S}_n(f_k) = (\hat{\sigma}_{n}[k])^2 \quad (k=1,\ldots,L),
\end{align}
is expected to be a better estimate of the PSDs than the periodogram computed with
$\hat{\mathbf{u}}$.

Intuitively, the proposed model is expected to accurately estimate smooth and block-sparse PSDs
by the following mechanism.
Since $\phi(\mathbf{v},\sigma)$ is basically
$\frac{\|\mathbf{v}\|^2}{2\sigma} + \frac{J}{2}\sigma$ (see \eqref{eq:defenition:phi}),
roughly speaking, we can consider that
the part \eqref{eq:PenaltyPhiPart} acts as
\begin{align*}
\lambda\sum_{n=1}^{N}\sum_{k = 1}^{L} 
\sum_{j=1}^{J}\left(\frac{|u_{j,n}[k]|^2}{2\sigma_n[k]} + \frac{\sigma_n[k]}{2}\right).
\end{align*}
Since $\sigma_n[k]$ estimates the square root of the PSDs,
\begin{align*}
\frac{|u_{j,n}[k]|^2}{2\sigma_n[k]}
\end{align*}
is expected to be an effective regularization
for $\mathbf{u}$
because the expectation of the squared magnitude of the realizations $\bar{\mathbf{u}}$ is close to the PSDs (see \eqref{eq:ExpectationPeriodogram} and Fig.~\ref{fig:illustPSDandPeriodogram}).
Refining $\mathbf{u}$ leads to an equally refined $\boldsymbol{\sigma}$
because $\boldsymbol{\sigma}$ is smoothed around the value in \eqref{eq:RelationSigmaToAvePerio}.
Thanks to these interactions, the proposed model is expected to effectively estimate
the frequency components and the PSDs simultaneously.

While $\|\mathbf{D}\boldsymbol{\sigma}_n \|_{1}$
with the first difference operator $\mathbf{D}$ in \eqref{eq:def:FirstDifferenceOp}
is a good choice for controlling the block structure,
more advanced smoothness priors can be incorporated to further improve the estimation accuracy of the PSDs.
Thanks to the nonnegativity of $\boldsymbol{\sigma}_n$,
many convex smoothness penalties designed for real-valued signals, such as
the high-order total variation \cite{Chan:HTV,Maso:HTV} that uses $\mathbf{D}^r$ $(r \geq 2)$ instead of $\mathbf{D}$ and the total generalized variation \cite{Bredies:TGB},
can be directly applied to $\boldsymbol{\sigma}_n$.
For instance, the proposed model with the high-order total variation
\begin{equation}
\label{eq:PropReg_GeneralDiffOp}
\left.\begin{aligned}
\minimize_{\mathbf{u} \in \mathbb{C}^{JNL}, \boldsymbol{\sigma} \in \mathbb{R}_{+}^{NL}}
\frac{1}{2}&\sum_{j=1}^{J}
\left\|\mathbf{y}_j-\sum_{n=1}^{N}\mathbf{A}_n\mathbf{G}\mathbf{u}_{j,n}\right\|^2\\
+&\lambda\sum_{n=1}^{N}\sum_{k = 1}^{L} \phi \left( (u_{j,n}[k])_{j=1}^{J},\sigma_n[k]\right)&\\
&\mbox{subject to }  \sum_{n=1}^{N}\|\mathbf{D}^r\boldsymbol{\sigma}_n \|_{1}  \leq \alpha
\end{aligned}\right\}
\end{equation}
can be solved similarly to the case of \eqref{eq:PropReg} by the ADMM-based algorithm
shown in Appendix \ref{appendix:OptimizationAlgorithm}.
In contrast, when these penalties are applied to, e.g., the magnitude of $\mathbf{u}$,
their convexity is lost (see, e.g., \cite{Ciuciu:SmoothPSD}),
which implies that a globally optimal solution is difficult to obtain.
Note that the application of these penalties to
$\mathbf{u}$, which is complex-valued, is not a suitable strategy
because the magnitude of $\mathbf{u}$ is smooth but
the phase of $\mathbf{u}$ is not smooth in most applications.

\section{Simulation Results}
\label{sect:experiment}
To demonstrate the effectiveness of the proposed approach,
we conduct numerical simulations on the PSD estimation for the PAWR shown in Example \ref{exmp:PAWR}.
Essentially, we follow the simulation setting in \cite{Yoshikawa:PAWR,Kitahara:PAWR_GRSS}.
Uniform elevation angles $\theta_1,\ldots,\theta_{N}$, ranging between $\ang{-15}$ and $\ang{30}$ degrees with $N=110$, are selected.
We synthesize the (discrete-time) PSD $S_n^{\star}(f)$ by
\begin{align*}
S_n^{\star}(f) = \frac{1}{T}\sum_{m=-\infty}^{\infty}G_n^{\star}\left(\frac{f-m}{T}\right)
\end{align*}
for each $n=1,\ldots,N$,
where $T$ is the pulse repetition time, and $G_n^{\star}(f)$ is a continuous-time Gaussian-shaped PSD
\begin{align*}
G_n^{\star}(f) = \frac{P_n}{\sqrt{2\pi}\varsigma_{n}}e^{-\frac{(f-\mu_{n})^2}{2\varsigma_n^2}},
\end{align*}
which is an appropriate model when, e.g., the atmospheric turbulence is dominant \cite{Bringi:BookDopplerRadar}.
The power $P_n$ is set from the actual reflection intensity measured by the PAWR at Osaka University on March 30, 2014.
We define the mean Doppler frequency $\mu_n$ by the certain sine curve used in \cite{Kitahara:PAWR_GRSS}.
The Doppler frequency width $\varsigma_n$ is converted from
the Doppler velocity width, which are chosen randomly from the uniform distribution of $[1,3]$ $\mathrm{[m/s]}$.
Note that this setting is more realistic than that presented in \cite{Kitahara:PAWR_GRSS}
where the Doppler velocity width is merely fixed to $2$ $\mathrm{[m/s]}$
at every elevation angle.
We set $X_n^\star[\ell]$ in \eqref{eq:def:random_process}
to the Gaussian process that has the specified PSD $S_n^{\star}(f)$,
and then generate its realizations $\bar{x}_{j,n}[\ell]\quad (\ell=1,\ldots,L)$
based on the probability distribution of $X_n^\star[\ell]\quad (\ell=1,\ldots,L)$,
which is computed in the way presented in \cite{Bringi:BookDopplerRadar,Kitahara:PAWR_GRSS}.
The observation vector $\mathbf{y}_{j}$ is given by \eqref{eq:ObservationPAWR},
where $\boldsymbol{\varepsilon}_j$ is generated as the white Gaussian noise of
the standard deviation $\sqrt{2.5}$.
The parameters of the PAWR are set as follows: $M=128$,
$\lambda_{\mathrm{cw}} = 31.8 \, [\mathrm{mm}]$,
$\Delta = 16.5 \, [\mathrm{mm}]$, and $T = 0.4 \, [\mathrm{ms}]$.
For the synthesis matrix $\mathbf{G}$ in
the observation model \eqref{eq:GeneralObservationModelFreqDom} in terms of the frequency components,
we test both $\mathbf{G}=\mathbf{F}^{\mathrm{H}}$ and
$\mathbf{G}=\mathbf{W}^{-1}\mathbf{F}^{\mathrm{H}}$
respectively for the standard DFT
in Example \ref{exmp:DFT} and
the windowed DFT in Example \ref{exmp:windowedDFT}.
We use the hamming window for the window function $\mathbf{w}$ in Example \ref{exmp:windowedDFT},
which is normalized to $\|\mathbf{w}\| = \sqrt{L}$, i.e., to the norm of the rectangular window $(1,1,\ldots,1)^{\top}$
\cite{Chandrasekar:BookDopplerRadar}.

We compare the proposed approach that jointly estimates the frequency components and the PSDs
with the existing approach
that first estimates the frequency components and subsequently the PSDs.
The proposed approach computes the estimate $\hat{S}_n(f_k)$
of the PSD by \eqref{eq:ProposedEstimatePSD}
with the solution $\hat{\boldsymbol{\sigma}}$ of the proposed model \eqref{eq:PropReg_GeneralDiffOp}
for the variable $\boldsymbol{\sigma}$.
For the frequency component estimators used in the existing approach,
we employ the mixed $\ell_2/\ell_1$ regularization model using fixed small-size overlapping blocks \cite{Kitahara:PAWR_GRSS},
which is state-of-the-art for the PAWR,
and the $\ell_1$ regularization model as a non-structured sparse model.
For the mixed $\ell_2/\ell_1$ regularization model,
we adopt the formulation based on latent group lasso \cite{Jacob:l12,Obozinski:LGL:TechRep},
which selects relevant blocks from the pre-defined overlapping blocks in the mixed $\ell_2/\ell_1$ norm, because its estimation accuracy is (slightly) better
than that of the simple overlapping blocks-based formulation in \cite{Kitahara:PAWR_GRSS}.
While the above nonlinear methods outperform the linear methods for the frequency component estimation  in \cite{Kitahara:PAWR_GRSS},
we also include the minimum mean square error (MMSE) beamformer \cite{Yoshikawa:PAWR},
which performs best among the linear methods in \cite{Kitahara:PAWR_GRSS}, for comparison.
Since the MMSE beamformer is a time-domain method that estimates
$\bar{\mathbf{x}}_{j,n}$ from \eqref{eq:def:ObservationModel_TimeDomain},
we compute the frequency components from the estimate of $\bar{\mathbf{x}}_{j,n}$
by using the DFT or the windowed DFT shown in Examples \ref{exmp:DFT} and \ref{exmp:windowedDFT} respectively.
From the estimated frequency components $\hat{\mathbf{u}}$,
the existing approach constructs
the estimate of the PSD
as the averaged periodogram
\begin{align}
\label{eq:def:EnsAvePerio}
\hat{S}_n^{\mathrm{(AP)}}(f_k) = \frac{1}{J}\sum_{j=1}^{J}|\hat{u}_{j,n}[k]|^2 \quad (k=1,\ldots,L).
\end{align}
We also test the post-smoothing for the existing approach.
Specifically, we employ the standard smoothing technique, i.e.,
\emph{Daniell method} \cite{Stoica:BookPSD,Brockwell:BookPSD}:
\begin{align*}
\hat{S}_n^{\mathrm{(SAP)}}(f_k) = \frac{1}{2R+1}\sum_{k'=k-R}^{k+R}\hat{S}_n^{\mathrm{(AP)}}(f_{k'})  \quad (k=1,\ldots,L),
\end{align*}
where $2R$ neighbor frequency bins\footnote{%
When $k' \notin \{1,\ldots,L\}$,
we instead use $k' - L\lfloor (k'-1)/L \rfloor$ because the anti-aliasing filter is not employed
(see also Example \ref{exmp:PAWR}).} are used for the smoothing.

\begin{table*}[!t]
%% increase table row spacing, adjust to taste
\renewcommand{\arraystretch}{1.5}
\footnotesize
\caption{A comparison of the methods in terms of the NMAE of the PSDs, where the result is averaged over $100$ independent simulations. Values shown in parenthesis for the existing methods are the NMAEs with the post-smoothing.}
\label{table:NMAE_PSD}
\centering
\begin{tabular}{l c c c c c}
\hline
Settings
& MMSE BF
& $\ell_{1}$ reg.
& Mixed $\ell_{2}/\ell_{1}$ reg.
& Proposed
\\
\hline
$L=32$, $J=1$, $\mathbf{G}=\mathbf{F}^{\mathrm{H}}$ (standard DFT in Example \ref{exmp:DFT})
& $1.1775$ $(0.9217)$
& $0.8771$ $(0.6802)$
& $0.7725$ $(0.6447)$
& $\mathbf{0.6278}$
\\
\hline
$L=32$, $J=1$, $\mathbf{G}=\mathbf{W}^{-1}\mathbf{F}^{\mathrm{H}}$ (windowed DFT in Example \ref{exmp:windowedDFT})
& $1.1055$ $(0.9494)$
& $0.8236$ $(0.7239)$
& $0.7644$ $(0.7004)$
& $\mathbf{0.6780}$
\\
\hline
$L=128$, $J=1$, $\mathbf{G}=\mathbf{F}^{\mathrm{H}}$ (standard DFT in Example \ref{exmp:DFT})
& $1.1808$ $(0.7396)$
& $0.8161$ $(0.4874)$
& $0.6843$ $(0.4906)$
& $\mathbf{0.4693}$
\\
\hline
$L=128$, $J=1$, $\mathbf{G}=\mathbf{W}^{-1}\mathbf{F}^{\mathrm{H}}$ (windowed DFT in Example \ref{exmp:windowedDFT})
& $1.1767$ $(0.7975)$
& $0.7899$ $(0.5428)$
& $0.6970$ $(0.5244)$
& $\mathbf{0.5097}$
\\
\hline
$L=32$, $J=2$, $\mathbf{G}=\mathbf{F}^{\mathrm{H}}$ (standard DFT in Example \ref{exmp:DFT})
& $0.9345$ $(0.7767)$
& $0.6917$ $(0.5582)$
& $0.6201$ $(0.5590)$
& $\mathbf{0.5580}$
\\
\hline
$L=32$, $J=2$, $\mathbf{G}=\mathbf{W}^{-1}\mathbf{F}^{\mathrm{H}}$ (windowed DFT in Example \ref{exmp:windowedDFT})
& $0.9292$ $(0.8318)$
& $0.6596$ $(0.5865)$
& $0.6169$ $(0.5790)$
& $\mathbf{0.5718}$
\\
\hline
$L=128$, $J=2$, $\mathbf{G}=\mathbf{F}^{\mathrm{H}}$ (standard DFT in Example \ref{exmp:DFT})
& $0.9815$ $(0.7057)$
& $0.6676$ $(0.4345)$
& $0.5700$ $(0.4623)$
& $\mathbf{0.4335}$
\\
\hline
$L=128$, $J=2$, $\mathbf{G}=\mathbf{W}^{-1}\mathbf{F}^{\mathrm{H}}$ (windowed DFT in Example \ref{exmp:windowedDFT})
& $0.9763$ $(0.7418)$
& $0.6435$ $(0.4684)$
& $0.5723$ $(0.4771)$
& $\mathbf{0.4548}$
\\
\hline
\end{tabular}
\end{table*}

\begin{table*}[!t]
%% increase table row spacing, adjust to taste
\renewcommand{\arraystretch}{1.5}
\footnotesize
\caption{Specific settings of the tuning parameters,
where $\varsigma_{\varepsilon}$ is the standard deviation of the noise,  $R$ is the parameter for the post-smoothing, $\lambda$ is the weight for the (block-)sparsity, $B$ is the block-size for the mixed $\ell_{2}/\ell_{1}$ regularization model, and $\alpha$ is the smoothing parameter for the proposed model.}
\label{table:setting_tuningParam}
\centering
\begin{tabular}{l c c c c c}
\hline
Settings
& MMSE BF
& $\ell_{1}$ reg.
& Mixed $\ell_{2}/\ell_{1}$ reg.
& Proposed
\\
\hline
$L=32$, $J=1$, $\mathbf{G}=\mathbf{F}^{\mathrm{H}}$
& $\varsigma_{\varepsilon} = \sqrt{2.5}$, $R=1$
& $\lambda = 0.003\frac{M}{N}$, $R=1$
& $\lambda = 0.03\frac{M}{N}$, $B = 7$, $R = 1$
& $\lambda = 0.02\frac{M}{N}$, $\alpha = 60N$
\\
\hline
$L=32$, $J=1$, $\mathbf{G}=\mathbf{W}^{-1}\mathbf{F}^{\mathrm{H}}$
& $\varsigma_{\varepsilon} = \sqrt{2.5}$, $R=1$
& $\lambda = 0.003\frac{M}{N}$, $R=1$
& $\lambda = 0.02\frac{M}{N}$, $B = 9$, $R = 1$
& $\lambda = 0.01\frac{M}{N}$, $\alpha = 60N$
\\
\hline
$L=128$, $J=1$, $\mathbf{G}=\mathbf{F}^{\mathrm{H}}$
& $\varsigma_{\varepsilon} = \sqrt{2.5}$, $R=5$
&$\lambda = 0.005\frac{M}{N}$, $R=4$
& $\lambda = 0.05\frac{M}{N}$, $B=36$, $R=4$
& $\lambda = 0.05\frac{M}{N}$, $\alpha = 20N$
\\
\hline
$L=128$, $J=1$, $\mathbf{G}=\mathbf{W}^{-1}\mathbf{F}^{\mathrm{H}}$
& $\varsigma_{\varepsilon} = \sqrt{2.5}$, $R=5$
& $\lambda = 0.003\frac{M}{N}$, $R=5$
& $\lambda = 0.02\frac{M}{N}$, $B=36$, $R=4$
& $\lambda = 0.03\frac{M}{N}$, $\alpha = 20N$
\\
\hline
$L=32$, $J=2$, $\mathbf{G}=\mathbf{F}^{\mathrm{H}}$
& $\varsigma_{\varepsilon} = \sqrt{2.5}$, $R=1$
& $\lambda = 0.01\frac{M}{N}$, $R=1$
& $\lambda = 0.02\frac{M}{N}$, $B = 7$, $R = 1$
& $\lambda = 0.03\frac{M}{N}$, $\alpha = 60N$
\\
\hline
$L=32$, $J=2$, $\mathbf{G}=\mathbf{W}^{-1}\mathbf{F}^{\mathrm{H}}$
& $\varsigma_{\varepsilon} = \sqrt{2.5}$, $R=1$
& $\lambda = 0.006\frac{M}{N}$, $R=1$
& $\lambda = 0.01\frac{M}{N}$, $B = 6$, $R = 1$
& $\lambda = 0.02\frac{M}{N}$, $\alpha = 60N$
\\
\hline
$L=128$, $J=2$, $\mathbf{G}=\mathbf{F}^{\mathrm{H}}$
& $\varsigma_{\varepsilon} = \sqrt{2.5}$, $R=4$
&$\lambda = 0.005\frac{M}{N}$, $R=4$
& $\lambda = 0.02\frac{M}{N}$, $B=36$, $R=3$
& $\lambda = 0.04\frac{M}{N}$, $\alpha = 20N$
\\
\hline
$L=128$, $J=2$, $\mathbf{G}=\mathbf{W}^{-1}\mathbf{F}^{\mathrm{H}}$
& $\varsigma_{\varepsilon} = \sqrt{2.5}$, $R=4$
&$\lambda = 0.004\frac{M}{N}$, $R=4$
& $\lambda = 0.02\frac{M}{N}$, $B=36$, $R=3$
& $\lambda = 0.03\frac{M}{N}$, $\alpha = 20N$
\\
\hline
\end{tabular}
\end{table*}

Table \ref{table:NMAE_PSD} shows the normalized mean absolute error (NMAE)
\begin{align*}
\frac{\sum_{n=1}^{N}\sum_{k=1}^{L}\left|S_n^{\star}(f_k) - \hat{S}_n(f_k)\right|}{\sum_{n=1}^{N}\sum_{k=1}^{L}S_n^{\star}(f_k)},
\end{align*}
which is averaged over $100$ independent simulations.
The tuning parameters of the methods are adjusted in the way that the best NMAE is obtained for each method and setting.
Table \ref{table:setting_tuningParam} shows
specific settings of the tuning parameters:
$\lambda$ for the importance of the (block)-sparsity,
$\alpha$ for the importance of the smoothness in the proposed model, the block-size $B$ for the mixed $\ell_2/\ell_1$ regularization model, $R$ for the post-smoothing, and
the standard deviation $\varsigma_{\varepsilon}$ of the noise for the MMSE beamformer.
Note that the MMSE beamformer uses the actual standard deviation $\sqrt{2.5}$ in the experiments
to achieve the best accuracy.
We simply set $r = 2$ in the proposed model, although the tuning of $r$ could improve the estimation accuracy.
While the case of $J=1$ is of particular interest in weather radar applications
to keep the frequency resolution of the PSDs (see Remark \ref{rmrk:MultipleObservation}),
we also show the results when $J$ is increased to $2$,
so as to elaborate on the effect of $J$.
In Table \ref{table:NMAE_PSD},
the proposed model is shown to achieve the best estimation accuracy for all the settings.
The post-smoothing is found to improve the estimation accuracies of the existing models;
however, their accuracies remain inferior to those of the proposed model.
While the NMSEs of the proposed model and the the existing sparse estimation models
combined with the post-smoothing are close for several cases when $J$ is increased to $2$,
the proposed model yields moderate improvements against them
for the cases of $J = 1$. Since the original frequency resolution is preserved
when $J = 1$ (see Remark \ref{rmrk:MultipleObservation}),
the proposed model has an advantage that it estimates PSDs accurately
without sacrificing the frequency resolution.
Note that the proposed model also has an advantage
that it has fewer tuning parameters than the the mixed $\ell_2/\ell_1$ regularization model with the post-smoothing
(see Table \ref{table:setting_tuningParam}).

\begin{figure*}[t]
  \centering
    \includegraphics[width=2.03\columnwidth]{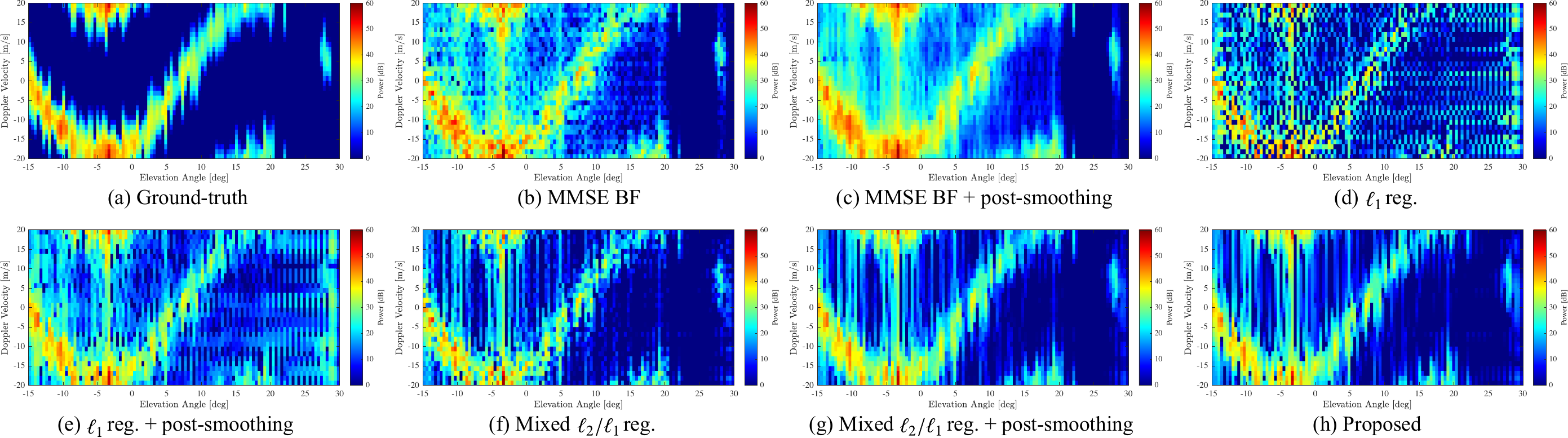}
  \caption{Ground-truth of the PSDs and their estimates for a simulation of the PAWR using the following settings: $L=32$, $J=1$, $\mathbf{G}=\mathbf{F}^{\mathrm{H}}$.}
  \label{fig:PSD_Exp_DFT_L=32_J=1}
\end{figure*}

\begin{figure*}[!h]
  \centering
    \includegraphics[width=2.03\columnwidth]{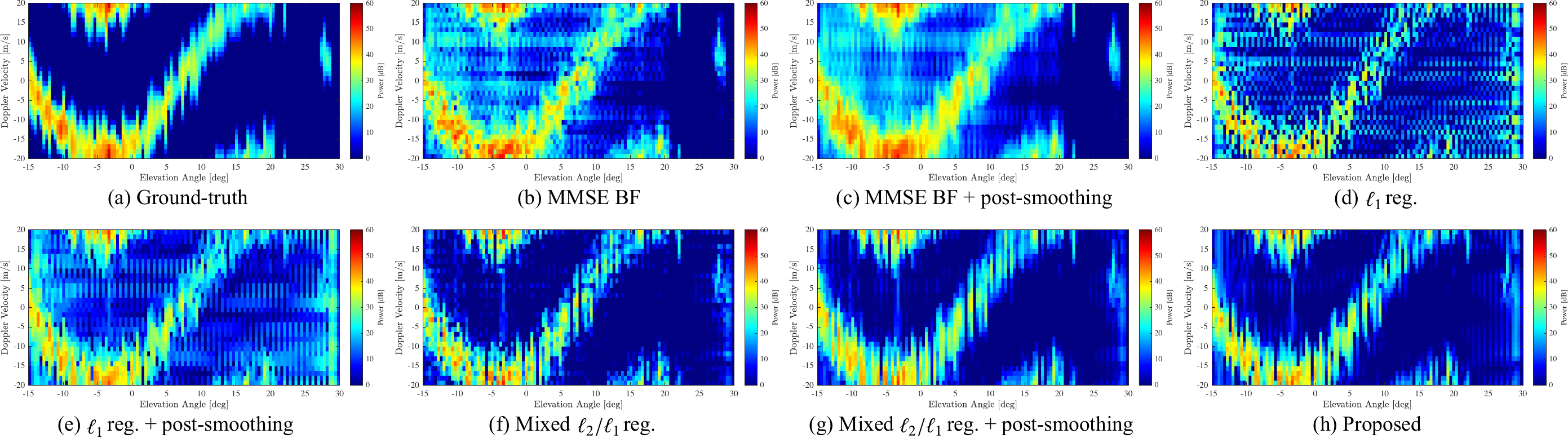}
  \caption{Ground-truth of the PSDs and their estimates for a simulation of the PAWR using the following settings: $L=32$, $J=1$, $\mathbf{G}=\mathbf{W}^{-1}\mathbf{F}^{\mathrm{H}}$.}
  \label{fig:PSD_Exp_WDFT_L=32_J=1}
\end{figure*}

\begin{figure*}[!h]
  \centering
    \includegraphics[width=2.03\columnwidth]{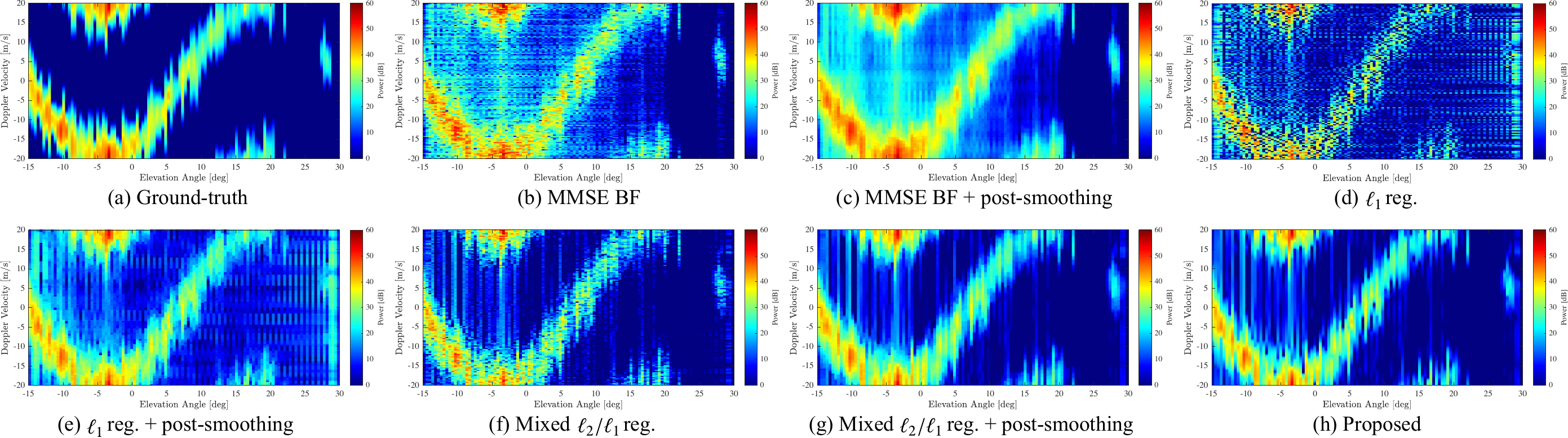}
  \caption{Ground-truth of the PSDs and their estimates for a simulation of the PAWR using the following settings: $L=128$, $J=1$, $\mathbf{G}=\mathbf{F}^{\mathrm{H}}$.}
  \label{fig:PSD_Exp_DFT_L=128_J=1}
\end{figure*}

\begin{figure*}[!h]
  \centering
    \includegraphics[width=2.03\columnwidth]{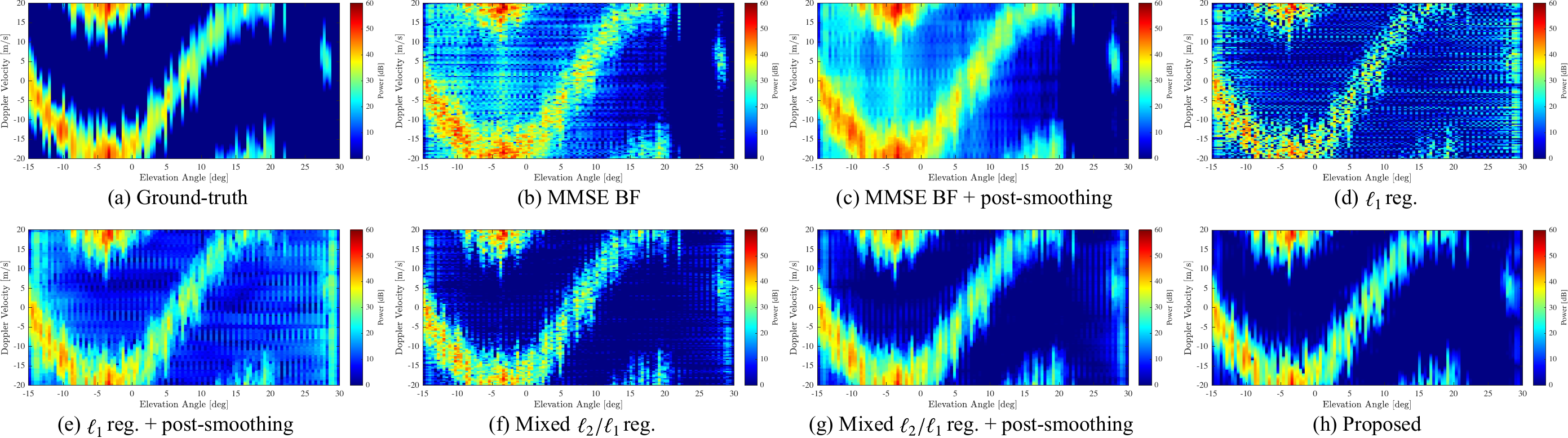}
  \caption{Ground-truth of the PSDs and their estimates for a simulation of the PAWR using the following settings: $L=128$, $J=1$, $\mathbf{G}=\mathbf{W}^{-1}\mathbf{F}^{\mathrm{H}}$.}
  \label{fig:PSD_Exp_WDFT_L=128_J=1}
\end{figure*}

\begin{figure*}[t]
  \centering
    \includegraphics[width=2.03\columnwidth]{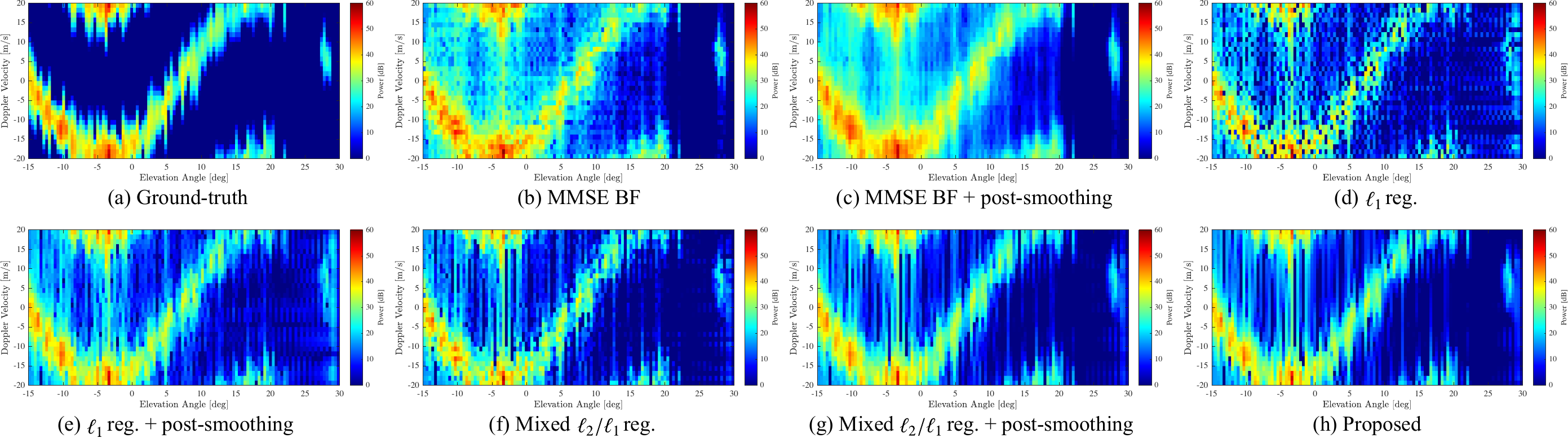}
  \caption{Ground-truth of the PSDs and their estimates for a simulation of the PAWR using the following settings: $L=32$, $J=2$, $\mathbf{G}=\mathbf{F}^{\mathrm{H}}$.}
  \label{fig:PSD_Exp_DFT_L=32_J=2}
\end{figure*}

\begin{figure*}[!h]
  \centering
    \includegraphics[width=2.03\columnwidth]{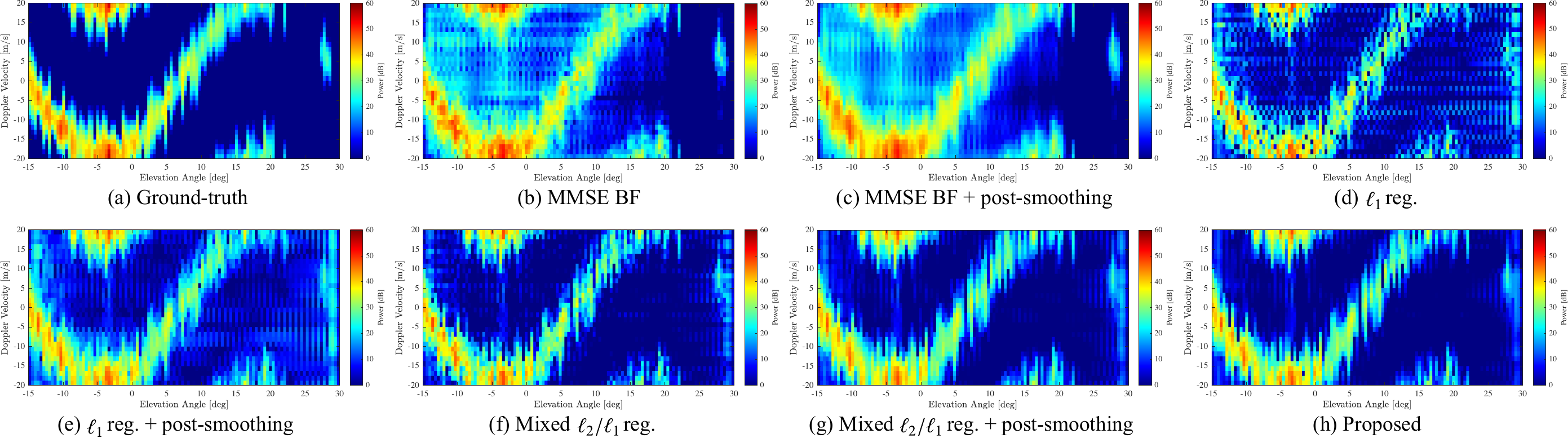}
  \caption{Ground-truth of the PSDs and their estimates for a simulation of the PAWR using the following settings: $L=32$, $J=2$, $\mathbf{G}=\mathbf{W}^{-1}\mathbf{F}^{\mathrm{H}}$.}
  \label{fig:PSD_Exp_WDFT_L=32_J=2}
\end{figure*}

\begin{figure*}[!h]
  \centering
    \includegraphics[width=2.03\columnwidth]{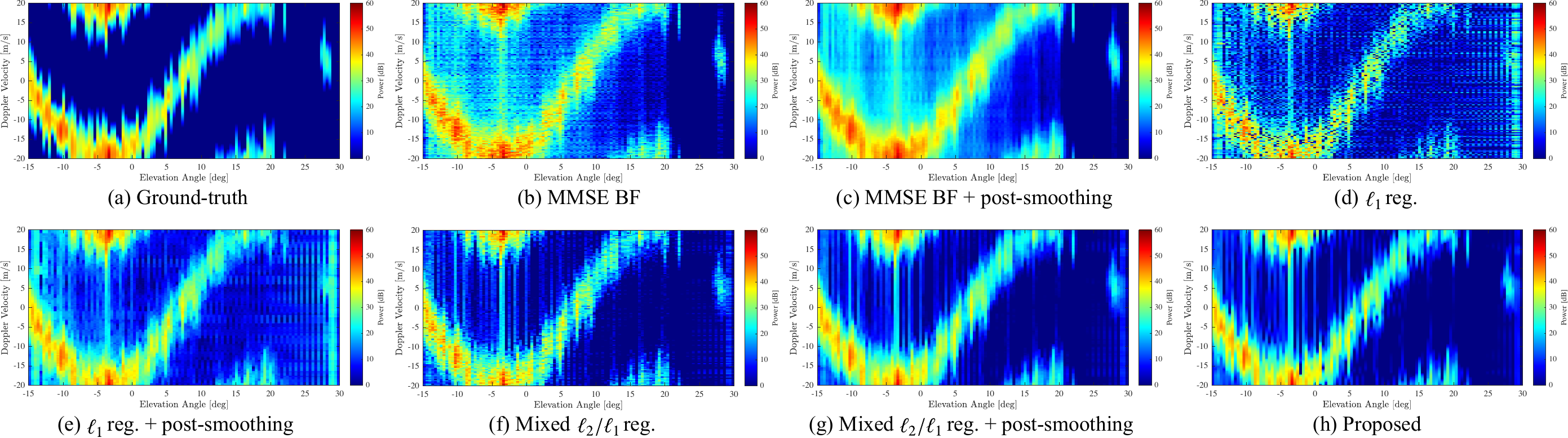}
  \caption{Ground-truth of the PSDs and their estimates for a simulation of the PAWR using the following settings: $L=128$, $J=2$, $\mathbf{G}=\mathbf{F}^{\mathrm{H}}$.}
  \label{fig:PSD_Exp_DFT_L=128_J=2}
\end{figure*}

\begin{figure*}[!h]
  \centering
    \includegraphics[width=2.03\columnwidth]{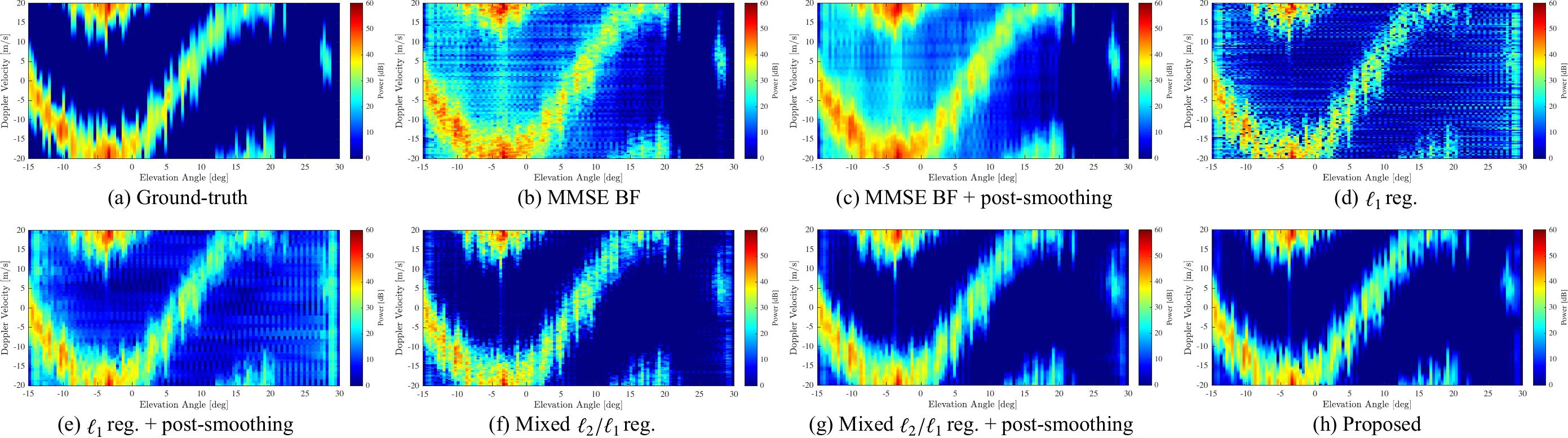}
  \caption{Ground-truth of the PSDs and their estimates for a simulation of the PAWR using the following settings: $L=128$, $J=2$, $\mathbf{G}=\mathbf{W}^{-1}\mathbf{F}^{\mathrm{H}}$.}
  \label{fig:PSD_Exp_WDFT_L=128_J=2}
\end{figure*}

\begin{figure*}[!h]
  \centering
    \includegraphics[width=2.1\columnwidth]{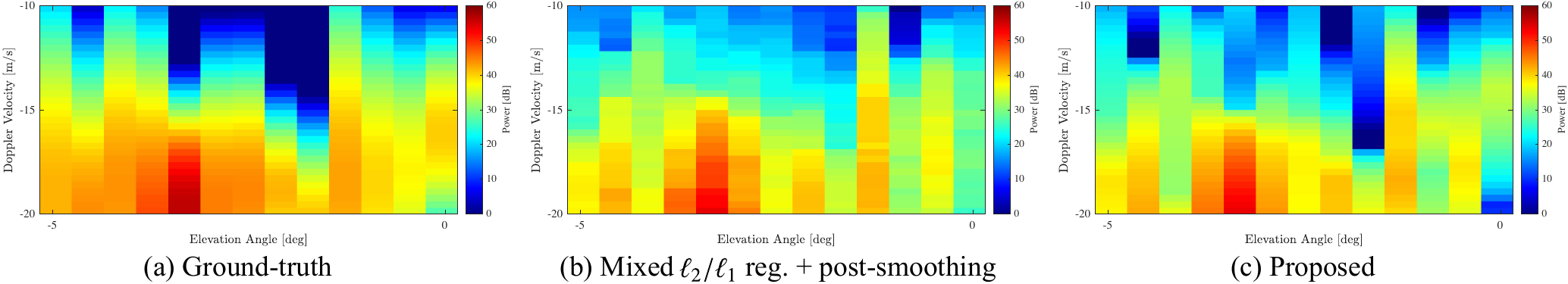}
  \caption{An enlarged view of Ground-truth of the PSDs and their estimates for the following settings: $L=128$, $J=1$, $\mathbf{G}=\mathbf{F}^{\mathrm{H}}$.}
  \label{fig:zoomed_PSD_Exp_DFT_L=128_J=1}
\end{figure*}

\begin{figure*}[!h]
  \centering
    \includegraphics[width=2.1\columnwidth]{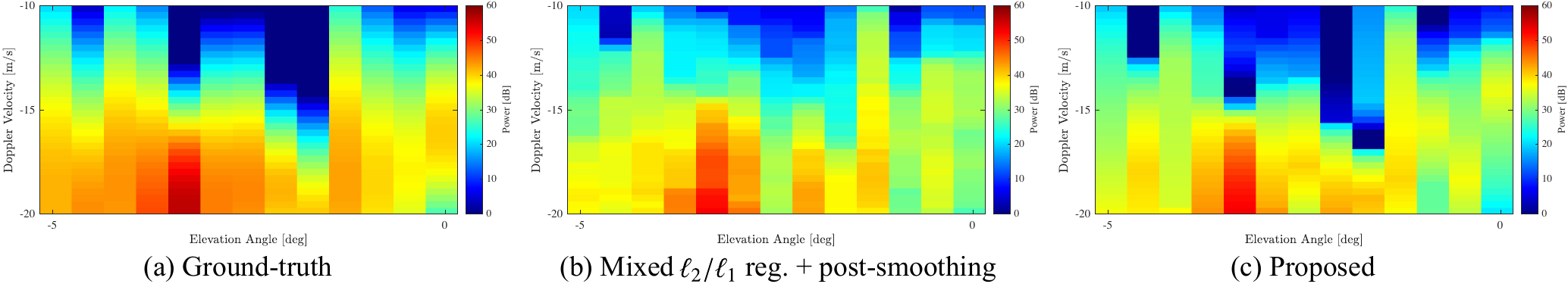}
  \caption{An enlarged view of Ground-truth of the PSDs and their estimates for the following settings: $L=128$, $J=1$, $\mathbf{G}=\mathbf{W}^{-1}\mathbf{F}^{\mathrm{H}}$.}
  \label{fig:zoomed_PSD_Exp_WDFT_L=128_J=1}
\end{figure*}

We show the ground-truth and the estimates of the PSDs for examples of simulations:
Fig.~\ref{fig:PSD_Exp_DFT_L=32_J=1} for $L=32$, $J=1$, $\mathbf{G}=\mathbf{F}^{\mathrm{H}}$,
Fig.~\ref{fig:PSD_Exp_WDFT_L=32_J=1} for $L=32$, $J=1$, $\mathbf{G}=\mathbf{W}^{-1}\mathbf{F}^{\mathrm{H}}$,
Fig.~\ref{fig:PSD_Exp_DFT_L=128_J=1} for $L=128$, $J=1$, $\mathbf{G}=\mathbf{F}^{\mathrm{H}}$,
Fig.~\ref{fig:PSD_Exp_WDFT_L=128_J=1} for $L=128$, $J=1$, $\mathbf{G}=\mathbf{W}^{-1}\mathbf{F}^{\mathrm{H}}$,
Fig.~\ref{fig:PSD_Exp_DFT_L=32_J=2} for $L=32$, $J=2$, $\mathbf{G}=\mathbf{F}^{\mathrm{H}}$,
Fig.~\ref{fig:PSD_Exp_WDFT_L=32_J=2} for $L=32$, $J=2$, $\mathbf{G}=\mathbf{W}^{-1}\mathbf{F}^{\mathrm{H}}$,
Fig.~\ref{fig:PSD_Exp_DFT_L=128_J=2} for $L=128$, $J=2$, $\mathbf{G}=\mathbf{F}^{\mathrm{H}}$,
Fig.~\ref{fig:PSD_Exp_WDFT_L=128_J=2} for $L=128$, $J=2$, $\mathbf{G}=\mathbf{W}^{-1}\mathbf{F}^{\mathrm{H}}$.
It can be seen from
Figs.~\ref{fig:PSD_Exp_DFT_L=32_J=1}–\ref{fig:PSD_Exp_WDFT_L=128_J=2}(b)(d)(f) that
the estimates of the existing models exhibit erratic oscillation as they do not exploit the smoothness of the PSDs.
In Figs.~\ref{fig:PSD_Exp_DFT_L=32_J=2}–\ref{fig:PSD_Exp_WDFT_L=128_J=2}(b)(d)(f) where $J$ is increased to $2$,
the erratic oscillation is slightly reduced but still clearly visible,
suggesting the limitation of the ensemble average \eqref{eq:def:EnsAvePerio} when $J$ is small (see Remark \ref{rmrk:MultipleObservation} for the reason why $J$ is set to be small in weather radar applications).
The post-smoothing is found to reduce the erratic oscillation to a certain extent, as shown
in Figs.~\ref{fig:PSD_Exp_DFT_L=32_J=1}–\ref{fig:PSD_Exp_WDFT_L=128_J=2}(c)(e)(g).
However, the sparsity of the estimate is impaired, i.e., the number of entries of large magnitude that are not present in the ground-truth increases, because the sparsity is not considered in the post-smoothing step.
In contrast, in Figs.~\ref{fig:PSD_Exp_DFT_L=32_J=1}–\ref{fig:PSD_Exp_WDFT_L=128_J=2}(h),
the proposed approach obtains the estimates that have both sparsity and smoothness.
While the estimates of the mixed $\ell_2/\ell_1$ regularization model after the post-smoothing
seem similar to the ground-truth at a glance of Figs.~\ref{fig:PSD_Exp_DFT_L=32_J=1}–\ref{fig:PSD_Exp_WDFT_L=128_J=2}(g),
erroneous spread of the nonzero components are more clearly seen
from enlarged views shown in Figs.~\ref{fig:zoomed_PSD_Exp_DFT_L=128_J=1} and \ref{fig:zoomed_PSD_Exp_WDFT_L=128_J=1}.
From Figs.~\ref{fig:zoomed_PSD_Exp_DFT_L=128_J=1} and \ref{fig:zoomed_PSD_Exp_WDFT_L=128_J=1}, we also see that
the proposed model estimates the area of nonzero components more accurately
than the mixed $\ell_2/\ell_1$ regularization model with the post-smoothing.
Since the area of the nonzero components is related to the existence of the corresponding wind velocity components (see Example \ref{exmp:PAWR}),
this is a significant advantage of the proposed approach for weather radar applications.

We also see that the erratic oscillation is still visible for the estimates of the MMSE beamformer
and the $\ell_1$ regularization model after the post-smoothing in Figs.~\ref{fig:PSD_Exp_DFT_L=32_J=1}–\ref{fig:PSD_Exp_WDFT_L=128_J=2}(c)(e), suggesting the limitation of the post-smoothing.
In particular, while the objective accuracies of the $\ell_1$ regularization model
after the post-smoothing are close to those of the proposed model when $J$ is increased to $2$ and the standard DFT is used,
the erratic oscillation is not eliminated as seen in Figs.~\ref{fig:PSD_Exp_DFT_L=32_J=2}(e) and \ref{fig:PSD_Exp_DFT_L=128_J=2}(e).
Compared to the $\ell_1$ regularization model,
the erratic oscillation is considerably reduced in the estimates of the mixed $\ell_2/\ell_1$ regularization model after the post-smoothing, as shown in Figs.~\ref{fig:PSD_Exp_DFT_L=32_J=1}–\ref{fig:PSD_Exp_WDFT_L=128_J=2}(g).
This could be attributed to the property that the block-sparse model does not necessarily promote the smoothness but force the components in the same block
to zeros together, which would be beneficial to promote the smoothness in the post-processing step.
Although the estimates of the mixed $\ell_2/\ell_1$ regularization model after the post-smoothing
seem smooth enough,
from enlarged views in Figs.~\ref{fig:zoomed_PSD_Exp_DFT_L=128_J=1} and \ref{fig:zoomed_PSD_Exp_WDFT_L=128_J=1},
we see that they have slight unnatural fluctuation,
and the estimates of the proposed model are smoother even compared with them.

The line-like artifacts in the estimates shown in Figs.~\ref{fig:PSD_Exp_DFT_L=32_J=1}, \ref{fig:PSD_Exp_DFT_L=128_J=1},
\ref{fig:PSD_Exp_DFT_L=32_J=2}, and \ref{fig:PSD_Exp_DFT_L=128_J=2}
are more or less reduced
in the estimates shown in Figs.~\ref{fig:PSD_Exp_WDFT_L=32_J=1}, \ref{fig:PSD_Exp_WDFT_L=128_J=1},
\ref{fig:PSD_Exp_WDFT_L=32_J=2}, and \ref{fig:PSD_Exp_WDFT_L=128_J=2}
thanks to the window function that reduces the heights of the sidelobes.
Although the estimates obtained with the windowed DFT are visually closer to the ground-truth than those obtained
with the standard DFT,
the objective accuracy shown in Table \ref{table:NMAE_PSD}
is not always improved perhaps because the window function increases the width of the mainlobe.
Since the line-like artifacts are caused due to the sidelobes
of the window function with finite $L$, the line-like artifacts are more reduced when $L=128$.
In particular, the line-like artifacts are almost completely eliminated
in the proposed estimates when $L=128$, as shown in
Figs.~\ref{fig:PSD_Exp_WDFT_L=128_J=1}(h) and \ref{fig:PSD_Exp_WDFT_L=128_J=2}(h).

\section{Conclusion}
\label{sect:Conclusion}
We presented a convex optimization model for the estimation of sparse and smooth PSDs of complex-valued random processes from noisy mixtures of realizations.
While the PSDs are related to the expectation of the magnitude of the frequency components of the realizations,
it has been difficult to exploit the smoothness of the PSDs
as naive penalties for the difference of the magnitude of the frequency components induce hard nonconvex optimization problems.
To resolve this difficulty, we designed the proposed model that jointly estimates complex-valued frequency components and nonnegative PSDs.
More precisely,
we first applied the optimally structured block-sparse model of \cite{Kuroda:BlockSparse}
for the frequency component estimation.
Then, to estimate the PSDs,
we newly leveraged the latent variable of the model,
which was originally introduced to optimize the block structure.
Namely, we demonstrate that the latent variable is
in fact related to the square root of the PSDs,
enabling us to exploit the smoothness of the PSDs via convex optimization
by penalizing the difference of the nonnegative latent variable.
Moreover,
to further enhance the smoothness of the PSDs of complex-valued random processes,
the proposed framework
can readily incorporate
many smoothness priors designed for real-valued signals.
Numerical experiments on the PAWR showed that the proposed approach achieved
better objective accuracy
and yielded visually better estimates compared with the existing sparse estimation models, even when they are combined with the post-smoothing.

\appendices

\begin{algorithm*}[t]
\DontPrintSemicolon
\setstretch{1.6}
\caption{Iterative solver for the proposed regularization model \eqref{eq:PropReg_GeneralDiffOp}}
\label{alg:OptimSolver}
\KwIn{$\gamma > 0$, $\mathbf{A} := [\mathbf{A}_1,\ldots,\mathbf{A}_N] \in \mathbb{C}^{d \times NL}$,
$\mathbf{x}_{j,n}^{(0)} \in \mathbb{C}^{L}$,
$\tilde{\mathbf{u}}_{j,n}^{(0)} \in \mathbb{C}^{L}$,
$\tilde{\boldsymbol{\sigma}}_{n}^{(0)} \in \mathbb{R}_{+}^{L}$,
$\boldsymbol{\eta}_{n}^{(0)} \in \mathbb{R}^{L}$,
$\mathbf{q}_{j,n}^{(0)}\in \mathbb{C}^{L}$,
$\mathbf{r}_{j,n}^{(0)}\in \mathbb{C}^{L}$,
$\boldsymbol{\tau}_{n}^{(0)} \in \mathbb{R}^{L}$,
$\boldsymbol{\zeta}_{n}^{(0)} \in \mathbb{R}^{L}$ $(j \in \{1,2,\ldots,J\} \mbox{ and } n \in \{1,2,\ldots,N\})$.}
\For{$i = 0,1,2,\ldots$}{
		\For{$n \in \{1,2,\ldots,N\}$}{
			\For{$j \in \{1,2,\ldots,J\}$}{
			$\mathbf{u}_{j,n}^{(i+1)} = (\mathbf{I}_{L} + \mathbf{G}^{\mathrm{H}}\mathbf{G})^{-1}
			\left(\mathbf{G}^{\mathrm{H}}
			(\mathbf{x}_{j,n}^{(i)}-\mathbf{q}_{j,n}^{(i)})+\tilde{\mathbf{u}}_{j,n}^{(i)}-\mathbf{r}_{j,n}^{(i)} \right)$
			}
			$\boldsymbol{\sigma}_{n}^{(i+1)} = (\mathbf{I}_{L} + (\mathbf{D}^{r})^{\top}\mathbf{D}^{r})^{-1}
			\left((\mathbf{D}^{r})^{\top}
			(\boldsymbol{\eta}_{n}^{(i)}-\boldsymbol{\zeta}_{n}^{(i)})+\tilde{\boldsymbol{\sigma}}_{n}^{(i)}-\boldsymbol{\tau}_{n}^{(i)} \right)$
		}
		\For{$j \in \{1,2,\ldots,J\}$}{
		$\left(\mathbf{x}_{j,n}^{(i+1)}\right)_{n=1}^{N}=
		(\mathbf{I}_{NL} + \gamma \mathbf{A}^{\mathrm{H}}\mathbf{A})^{-1}
		\left(\gamma\mathbf{A}^{\mathrm{H}}\mathbf{y}_j + 
		\left(\mathbf{G}\mathbf{u}_{j,n}^{(i+1)}+\mathbf{q}_{j,n}^{(i)}\right)_{n=1}^{N}\right)$
		}
		\For{$(n,k) \in \{1,2,\ldots,N\} \times \{1,2,\ldots,L \}$}{
			$\left( (\tilde{u}_{j,n}^{(i+1)}[k])_{j=1}^{J},\tilde{\sigma}_n^{(i+1)}[k]\right)=
			\mathrm{prox}_{\gamma\lambda\phi}
			\left( \left(u_{j,n}^{(i+1)}[k]+r_{j,n}^{(i)}[k]\right)_{j=1}^{J},\sigma_n^{(i+1)}[k]+\tau_n^{(i)}[k]\right)$
		}
		$\left(\boldsymbol{\eta}_{n}^{(i+1)}\right)_{n=1}^{N} = 
		P_{B_1^\alpha}\left( \left(\mathbf{D}^{r}\boldsymbol{\sigma}_{n}^{(i+1)} + \boldsymbol{\zeta}_{n}^{(i)}\right)_{n=1}^N \right)$\;
		\For{$n \in \{1,2,\ldots,N\}$}{
			\For{$j \in \{1,2,\ldots,J\}$}{
			$\mathbf{q}_{j,n}^{(i+1)} = \mathbf{q}_{j,n}^{(i)} + \mathbf{G}\mathbf{u}_{j,n}^{(i+1)} - \mathbf{x}_{j,n}^{(i+1)}$\;
			$\mathbf{r}_{j,n}^{(i+1)} = \mathbf{r}_{j,n}^{(i)} + \mathbf{u}_{j,n}^{(i+1)} - \tilde{\mathbf{u}}_{j,n}^{(i+1)}$
			}
		$\boldsymbol{\tau}_{n}^{(i+1)} = \boldsymbol{\tau}_{n}^{(i)} + \boldsymbol{\sigma}_{n}^{(i+1)} - \tilde{\boldsymbol{\sigma}}_{n}^{(i+1)}$\;
		$\boldsymbol{\zeta}_{n}^{(i+1)} = \boldsymbol{\zeta}_{n}^{(i)} + \mathbf{D}^{r}\boldsymbol{\sigma}_{n}^{(i+1)} - \boldsymbol{\eta}_{n}^{(i+1)}$
		}
		}
\end{algorithm*}

\section{\break Solver for Proposed Regularization Model}
\label{appendix:OptimizationAlgorithm}
The proposed regularization model \eqref{eq:PropReg_GeneralDiffOp}
can be solved by using the proximal splitting techniques \cite{Chambolle:PDS,Gabay:ADMM,Eckstein:ADMM,Combettes:ProxSplit,Condat:prox}
with the closed-form computation of the proximity operator of $\phi$.
As a concrete example, 
using the ADMM \cite{Gabay:ADMM,Eckstein:ADMM},
we provide an iterative solver that is guaranteed to converge to
an optimal solution of \eqref{eq:PropReg_GeneralDiffOp}.
The ADMM solves the following convex optimization problem
\begin{align}
\label{eq:OptimizationProblem_ADMMform}
\minimize_{v \in \mathcal{V}, w \in \mathcal{W}} F(v) + G(w) \mbox{ subject to }
w = \mathcal{L} v
\end{align}
by the iterations
\begin{equation*}
\left\lfloor \begin{aligned}
v^{(i+1)} & \in \arg\min_{v \in \mathcal{V}} \left[\gamma F(v) +
\frac{1}{2}\|w^{(i)}-z^{(i)}-\mathcal{L}v\|^2\right]\\
w^{(i+1)} &=\mathrm{prox}_{\gamma G}(\mathcal{L} v^{(i+1)} + z^{(i)})\\
z^{(i+1)} &= z^{(i)}+\mathcal{L}v^{(i+1)} - w^{(i+1)},
\end{aligned}\right.
\end{equation*}
where we suppose that
$\mathcal{V}$ and $\mathcal{W}$
are finite-dimensional Hilbert spaces,
$F$ and $G$ are proper lower semicontinuous convex functions,
$\mathcal{L}$ is a linear operator,
$\mathrm{prox}_{\gamma G}(w) := \mathrm{arg} \min_{\omega \in \mathcal{W}}
\left[\gamma G(\omega) + \frac{1}{2}\|w-\omega\|^2\right]$ is the proximity operator of
$\gamma G$, and $\gamma > 0$.

To apply the ADMM,
we rewrite \eqref{eq:PropReg_GeneralDiffOp} as
\begin{equation}
\label{eq:PropRegADMMform}
\left.\begin{aligned}
\minimize_{\mathbf{u},\boldsymbol{\sigma},\mathbf{x},\tilde{\mathbf{u}},\tilde{\boldsymbol{\sigma}},\boldsymbol{\eta}}
&F(\mathbf{u},\boldsymbol{\sigma})+G(\mathbf{x},\tilde{\mathbf{u}},\tilde{\boldsymbol{\sigma}},\boldsymbol{\eta})\\
\mbox{subject to }  &\mathbf{x}_{j,n} = \mathbf{G}\mathbf{u}_{j,n} \quad(\forall j,n) \\
&\tilde{\mathbf{u}}_{j,n} = \mathbf{u}_{j,n} \quad(\forall j,n) \\
&\tilde{\boldsymbol{\sigma}}_{n} = \boldsymbol{\sigma}_{n} \quad(\forall n) \\
&\boldsymbol{\eta}_{n} = \mathbf{D}^r\boldsymbol{\sigma}_{n} \quad(\forall n)
\end{aligned}\right\},
\end{equation}
where we let
\begin{align*}
F(\mathbf{u},\boldsymbol{\sigma}) &:= 0,\\
G(\mathbf{x},\tilde{\mathbf{u}},\tilde{\boldsymbol{\sigma}},\boldsymbol{\eta}) &:= \frac{1}{2}\sum_{j=1}^{J}
\left\|\mathbf{y}_j-\sum_{n=1}^{N}\mathbf{A}_n\mathbf{x}_{j,n}\right\|^2\\
+\lambda&\sum_{n=1}^{N}\sum_{k = 1}^{L} \phi \left( (\tilde{u}_{j,n}[k])_{j=1}^{J},\tilde{\sigma}_n[k]\right)+\iota_{B_1^{\alpha}}\left(\boldsymbol{\eta} \right),
\end{align*}
and $\iota_{B_1^{\alpha}}(\boldsymbol{\eta})$ is the indicator function of the $\ell_1$ ball, i.e.,
\begin{align*}
\iota_{B_1^{\alpha}}(\boldsymbol{\eta}) :=
\begin{dcases}
0, &\mbox{if }\sum_{n=1}^{N}\|\boldsymbol{\eta}_n\|_1 \leq \alpha;\\
\infty, &\mbox{otherwise}.
\end{dcases}
\end{align*}
Since the constraint of \eqref{eq:PropRegADMMform}
can be expressed in the form of \eqref{eq:OptimizationProblem_ADMMform},
we can therefore apply the ADMM to \eqref{eq:PropRegADMMform},
and obtain the iterative algorithm shown in Algorithm \ref{alg:OptimSolver}.
For our formulation,
since the minimizer of the first step of the ADMM is unique,
the convergence to an optimal solution of \eqref{eq:PropRegADMMform} follows from \cite{Condat:prox}.
From the equivalence between \eqref{eq:PropReg_GeneralDiffOp}
and \eqref{eq:PropRegADMMform},
$(\mathbf{u}^{(i)})_{i=1}^{\infty}$ and $(\boldsymbol{\sigma}^{(i)})_{i=1}^{\infty}$
generated by Algorithm \ref{alg:OptimSolver}
converges to the solution of \eqref{eq:PropReg_GeneralDiffOp}
for the variables $\mathbf{u}$ and $\boldsymbol{\sigma}$ respectively.

The operators in Algorithm \ref{alg:OptimSolver} can be computed as follows.
Expressing $\phi(\mathbf{v},\sigma)$ as the sum of
the perspective function \cite{Combettes:Perspective,Combettes:PerspectiveProx,Combettes:PerspectiveML}
of $\frac{\|\mathbf{v}\|^2}{2}$ and the linear function $\frac{J}{2}\sigma$,
based on \cite[Example 2.4]{Combettes:PerspectiveML},
we can compute $\mathrm{prox}_{\kappa \phi}$ for $\kappa = \gamma\lambda$ by
\begin{align}
\nonumber
&\mathrm{prox}_{\kappa \phi}(\mathbf{v},\sigma)\\
\label{eq:ProxSquareFuncWithWeight}
&=\begin{cases}
(\boldsymbol{0}, 0), &\mbox{if }2\kappa\sigma+\|\mathbf{v}\|^2 \leq J\kappa^2;\\
(\boldsymbol{0}, \sigma-\frac{\kappa J}{2}), &\mbox{if } \mathbf{v} = \boldsymbol{0} \mbox{ and }2\sigma > J\kappa;\\
\left(\mathbf{v} - \kappa s \frac{\mathbf{v}}{\|\mathbf{v}\|},\sigma+\kappa\frac{s^2-J}{2}\right), &\mbox{otherwise},
\end{cases}
\end{align}
where $s > 0$ is the unique positive root of 
\begin{align*}
s^3 + \left(\frac{2}{\kappa}\sigma+2-J\right)s-\frac{2}{\kappa}\|\mathbf{v}\| = 0,
\end{align*}
and can be explicitly given based on Cardano's formula as follows \cite{Kuroda:BlockSparse,Bauschke:RootCubic}.
Let $p = \frac{2}{\kappa}\sigma+2-J$ and $D = -\frac{\|\mathbf{v}\|^2}{\kappa^2}-\frac{p^3}{27}$. Then,
\begin{align*}
s = \begin{dcases}
\sqrt[3]{\frac{\|\mathbf{v}\|}{\kappa} + \sqrt{-D}} + \sqrt[3]{\frac{\|\mathbf{v}\|}{\kappa} - \sqrt{-D}}, &\mbox{if } D < 0;\\
2\sqrt[3]{\frac{\|\mathbf{v}\|}{\kappa}}, &\mbox{if } D = 0;\\
2\sqrt{-\frac{p}{3} } \cos\left(\frac{\arctan(\kappa\sqrt{D}/\|\mathbf{v}\|) }{3} \right), &\mbox{if } D > 0,
\end{dcases}
\end{align*}
where $\sqrt[3]{\cdot}$ denotes the real cubic root.
The $\ell_1$ ball projection $P_{B_1^\alpha}$, which is the proximity operator of $\iota_{B_1^{\alpha}}(\boldsymbol{\eta})$, can be computed in $\mathcal{O}(NL)$ expected complexity, e.g., by the algorithm of \cite{Condat:l1ballprojection}.
\subsection*{Implementation for weather radar applications}
The matrix inversions in Algorithm \ref{alg:OptimSolver}
can be efficiently computed
for application to weather radars as follows.
\begin{enumerate}
\item[a)] To efficiently compute the inversion of $(\mathbf{I}_{L} + \mathbf{G}^{\mathrm{H}}\mathbf{G})$, we use the property
\begin{align*}
\mathbf{G}\mathbf{G}^{\mathrm{H}} = \mathrm{diag}(\boldsymbol{\nu}) \in \mathbb{R}_{++}^{L},
\end{align*}
which holds for, e.g., the DFT in Example \ref{exmp:DFT} and the windowed DFT in Example \ref{exmp:windowedDFT}.
More precisely, from this property, we have
\begin{align*}
(\mathbf{I}_{L} + \mathbf{G}^{\mathrm{H}}\mathbf{G})^{-1}
&= \mathbf{I}_{L} - \mathbf{G}^{\mathrm{H}}(\mathbf{I}_{L} + \mathbf{G}\mathbf{G}^{\mathrm{H}})^{-1}\mathbf{G}\\
&= \mathbf{I}_{L} - \mathbf{G}^{\mathrm{H}}(\mathbf{I}_{L} + \mathrm{diag}(\boldsymbol{\nu}))^{-1}\mathbf{G}\\
&= \mathbf{I}_{L} - \mathbf{G}^{\mathrm{H}}\mathrm{diag}\left( \left(\frac{1}{1+\nu_{\ell}}\right)_{\ell=1}^{L}\right)\mathbf{G},
\end{align*}
where the first equality follows from the Sherman-Morrison-Woodbury matrix inversion lemma \cite{Golub:Matrix}.
Note that the multiplications of $\mathbf{G}$ and  $\mathbf{G}^{\mathrm{H}}$ can be computed in $\mathcal{O}(L \log L)$
by the fast Fourier transform (FFT) for Examples \ref{exmp:DFT} and \ref{exmp:windowedDFT}.
\item[b)] 
Since $\mathbf{D}$ in \eqref{eq:def:FirstDifferenceOp} is a circulant matrix, the inversion of $(\mathbf{I}_{L} + (\mathbf{D}^r)^{\top}\mathbf{D}^r)$ can be computed
in $\mathcal{O}(L \log L)$ with the FFT \cite{Golub:Matrix}.
\item[c)] 
For the PAWR shown in Example \ref{exmp:PAWR},
the inversion of $\mathbf{I}_{NL} + \mathbf{A}^{\mathrm{H}}\mathbf{A} \in \mathbb{C}^{NL \times NL}$
can be computed in $\mathcal{O}(N^3)$, independently of the value of $L$,
because it can be translated into
a block-diagonal matrix after some permutations.
We show this explicitly in another way. Notice that $(\mathbf{x}_{j,n}^{(i+1)})_{n=1}^{N}$ in Algorithm \ref{alg:OptimSolver}
is the solution of
\begin{equation}
\label{eq:def:StepForDataFidelity}
\left.\begin{aligned}
\minimize_{\left(\mathbf{x}_{j,n}\right)_{n=1}^{N}}
\frac{\gamma}{2}&\left\|\mathbf{y}_j-\sum_{n=1}^{N}\mathbf{A}_n\mathbf{x}_{j,n}\right\|^2\\
+ \frac{1}{2}&\sum_{n=1}^{N}\|\mathbf{G}\mathbf{u}_{j,n}^{(i+1)}+\mathbf{q}_{j,n}^{(i)}-\mathbf{x}_{j,n} \|^2
\end{aligned}\right\}.
\end{equation}
From the definitions of $\mathbf{y}_j^{\mathrm{(pawr)}}$ and $\mathbf{A}_n^{\mathrm{(pawr)}}$
in Example \ref{exmp:PAWR}, we have
\begin{align*}
&\left\|\mathbf{y}_j^{\mathrm{(pawr)}}-\sum_{n=1}^{N}\mathbf{A}_n^{\mathrm{(pawr)}}\mathbf{x}_{j,n}\right\|^2\\
=&\left\|
\begin{pmatrix}
\mathbf{y}_j[1]-\sum_{n=1}^{N}x_{j,n}[1]\mathbf{a}(\theta_n)\\
\vdots\\
\mathbf{y}_j[L]-\sum_{n=1}^{N}x_{j,n}[L]\mathbf{a}(\theta_n)
\end{pmatrix}
\right\|^2\\
=&\left\|
\begin{pmatrix}
\mathbf{y}_j[1]^{\top}-\sum_{n=1}^{N}x_{j,n}[1]\mathbf{a}(\theta_n)^{\top}\\
\vdots\\
\mathbf{y}_j[L]^{\top}-\sum_{n=1}^{N}x_{j,n}[L]\mathbf{a}(\theta_n)^{\top}
\end{pmatrix}
\right\|_{\mathrm{fro}}^2\\
=&\|\mathbf{Y}_j^{\top} - \mathbf{X}_j\mathbf{S}^{\top} \|_{\mathrm{fro}}^2,
\end{align*}
where we let
\begin{align*}
\mathbf{Y}_j :=& (\mathbf{y}_{j}[1],\ldots,\mathbf{y}_{j}[L]) \in \mathbb{C}^{M \times L},\\
\mathbf{X}_j :=& (\mathbf{x}_{j,1},\ldots,\mathbf{x}_{j,N}) \in \mathbb{C}^{L \times N},\\
\mathbf{S} :=& (\mathbf{a}(\theta_1),\ldots,\mathbf{a}(\theta_N)) \in \mathbb{C}^{M \times N},
\end{align*}
and $\|\cdot\|_{\mathrm{fro}}$ denotes the Frobenius norm of the matrix.
From this expression, it is clear that
the step \eqref{eq:def:StepForDataFidelity}
can be solved in $\mathcal{O}(N^3)$ for the inversion regarding $\mathbf{S}^{\top}$.
Namely, $(\mathbf{x}_{j,n}^{(i+1)})_{n=1}^{N}$ in Algorithm \ref{alg:OptimSolver}
is obtained by
\begin{align*}
\Bigl(&\mathbf{x}_{j,1}^{(i+1)},\ldots,\mathbf{x}_{j,N}^{(i+1)}\Bigr)\\
&=\left(\gamma \mathbf{Y}_j^{\top}\mathbf{S}^{*} + 
		\mathbf{G}\mathbf{U}_j^{(i+1)}+\mathbf{Q}_j^{(i)}\right)(\mathbf{I}_{N} + \gamma \mathbf{S}^{\top}\mathbf{S}^{*})^{-1},
\end{align*}
where $\mathbf{S}^{*}$ is the complex conjugate of $\mathbf{S}$, and
\begin{align*}
\mathbf{U}_j^{(i+1)} :=& \left(\mathbf{u}_{j,1}^{(i+1)},\ldots,\mathbf{u}_{j,N}^{(i+1)}\right) \in \mathbb{C}^{L \times N},\\
\mathbf{Q}_j^{(i)} :=& \left(\mathbf{q}_{j,1}^{(i)},\ldots,\mathbf{q}_{j,N}^{(i)}\right) \in \mathbb{C}^{L \times N}.
\end{align*}
\end{enumerate}
Meanwhile, we remark that the inversions in Algorithm \ref{alg:OptimSolver} are the same for all the iterations,
and thus can be computed in advance and stored in the memory.

\section*{Acknowledgment}
We would like to thank the anonymous reviewers for their valuable comments on the original version of the manuscript.

\bibliographystyle{IEEEtran}

\begin{IEEEbiography}[{\includegraphics[width=1in,height=1.25in,clip,keepaspectratio]{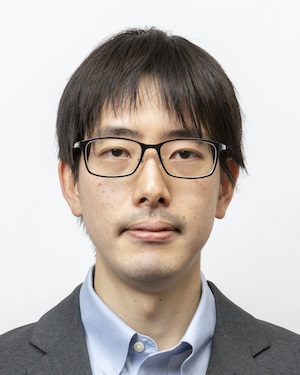}}]{Hiroki Kuroda}
(Member, IEEE) received the B.E. degree in computer science, and the M.E. and Ph.D. degrees in information and communications engineering from the Tokyo Institute of Technology, Tokyo, Japan, in 2013, 2015 and 2019, respectively.

In 2019, he was a Postdoctoral Researcher with the National Institute of Advanced Industrial Science and Technology, Ibaraki, Japan.
From 2020 to 2022, he was an Assistant Professor with the College of Information Science and Engineering, Ritsumeikan University, Shiga, Japan.
Since 2022, he has been an Assistant Professor with the Department of Information and Management Systems Engineering, Nagaoka University of Technology, Niigata, Japan. His research interests are in signal processing and its applications, inverse problem, sparse modeling, and optimization theory.

He was the recipient of the Young Researcher Award from the IEICE Technical Group on Signal Processing in 2018 and the Seiichi Tejima Doctoral Dissertation Award from the Tokyo Institute of Technology in 2020.
\end{IEEEbiography}

\begin{IEEEbiography}[{\includegraphics[width=1in,height=1.25in,clip,keepaspectratio]{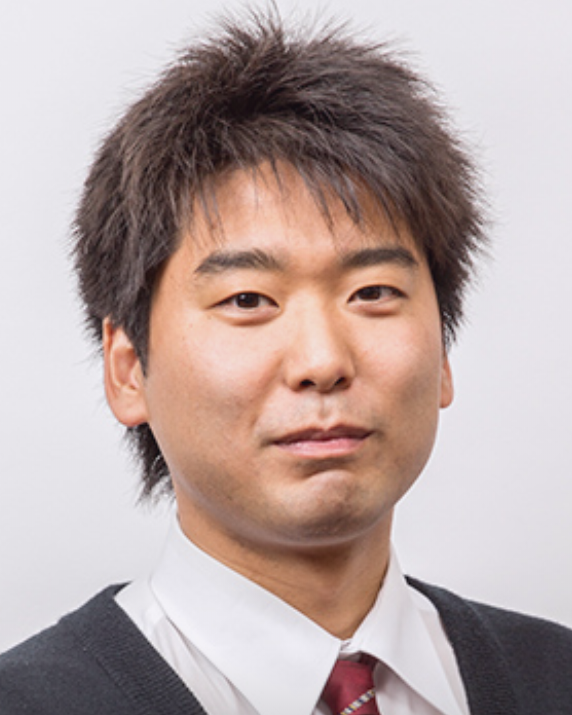}}]{Daichi Kitahara} (Member, IEEE) received the B.E. degree in computer science in 2012, and the M.E. and Ph.D. degrees in communications and computer engineering in 2014 and 2017 from the Tokyo Institute of Technology, Tokyo, Japan, respectively.

From 2017 to 2022, he was an Assistant Professor with the College of Information Science and Engineering, Ritsumeikan University, Shiga, Japan.
Currently he is a Researcher with the Graduate School of Engineering, Osaka University. 
His research interests are in signal processing and its applications, inverse problem, optimization theory, and multivariate spline theory.

He was the recipient of the Young Researcher Awards from the IEICE Technical Group on Signal Processing, the SICE Technical Committee on Remote Sensing, and the IEEE Computational Intelligence Society Japan Chapter in 2013, 2015, and 2016, respectively. He also was the recipient of the IEEE Signal Processing Society (SPS) Japan Student Journal Paper Award from the IEEE SPS Tokyo Joint Chapter in 2016 and the Yasujiro Niwa Outstanding Paper Award from the Tokyo Denki University in 2018.
\end{IEEEbiography}

\begin{IEEEbiography}[{\includegraphics[width=1in,height=1.25in,clip,keepaspectratio]{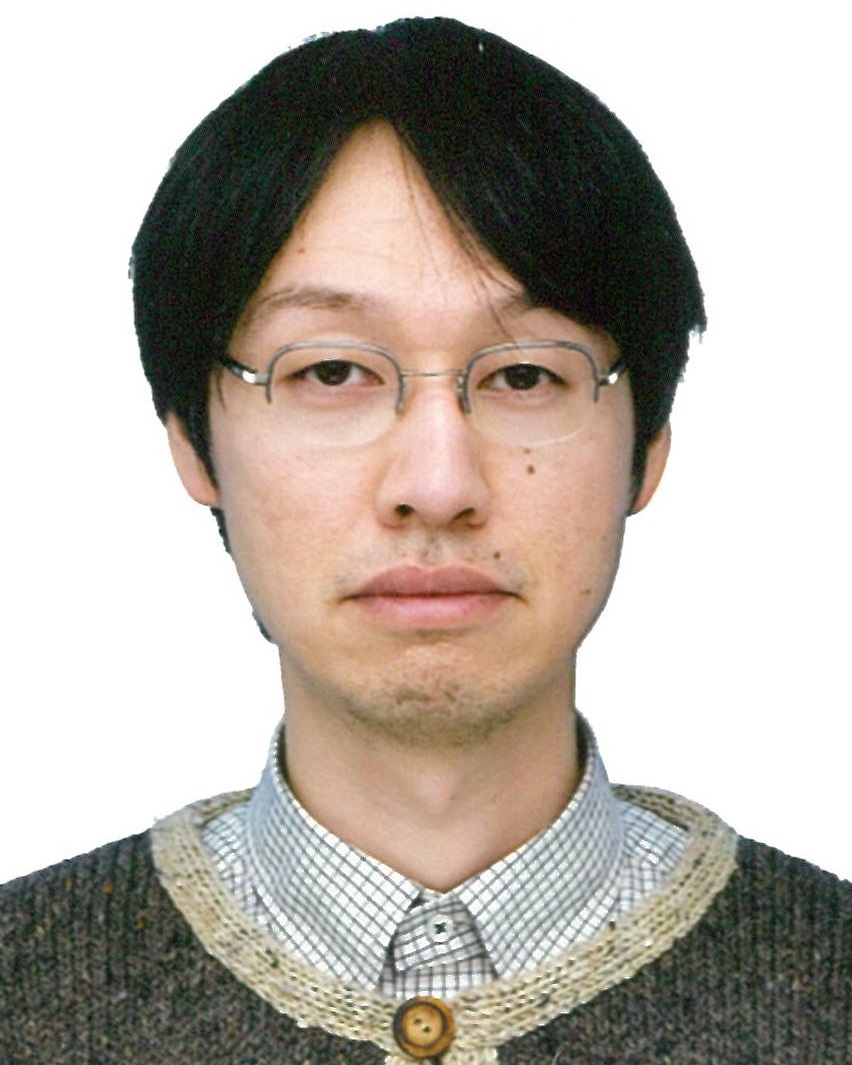}}]{Eiichi Yoshikawa} (Member, IEEE) received the B.E. degree in aerospace engineering from Osaka Prefecture University, Sakai, Japan, in 2005, and the M.E. and Ph.D. degrees from Osaka University, Suita, Japan, in 2008 and 2011, respectively.

In 2011, he was a Post-Doctoral Researcher with Osaka University and Colorado State University, Fort Collins, CO, USA, and a Post-Doctoral Fellow of the Research Fellowship for Young Scientists sponsored by the Japan Society for the Promotion of Science (JSPS). In 2012, he joined the Japan Aerospace Exploration Agency (JAXA), Mitaka, Tokyo, Japan, where he is currently an Associate Senior Researcher. He is also a Research Scientist with Colorado State University, Fort Collins, CO, USA. His research interests include weather radar remote sensing, radar-based analyses, and applications for general and aviation weather.
\end{IEEEbiography}

\begin{IEEEbiography}[{\includegraphics[width=1in,height=1.25in,clip,keepaspectratio]{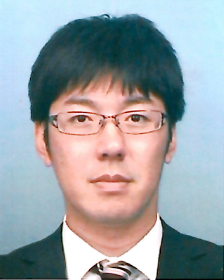}}]{Hiroshi Kikuchi} (Member, IEEE) received the B.S. degree from Department of Engineering, Doshisha University, Kyoto, Japan, in 2008 and the M.S., and Ph.D. degrees from the Division of Electrical, Electronic and Information Engineering, Osaka University, Suita, Japan, in 2010 and 2013, respectively.

He joined the Division of Electrical, Electric and Information Engineering, Osaka University as a Specially Appointed Researcher in 2013. In 2017, he was a Research Assistant Professor at Tokyo Metropolitan University. In 2018, he joined the University of Electro Communications, where he is currently an Associate Professor.
His research specialties are the remote sensing for an atmospheric electricity with space-borne platforms, and the weather radar remote sensing and a development of the radar system.

\end{IEEEbiography}

\begin{IEEEbiography}[{\includegraphics[width=1in,height=1.25in,clip,keepaspectratio]{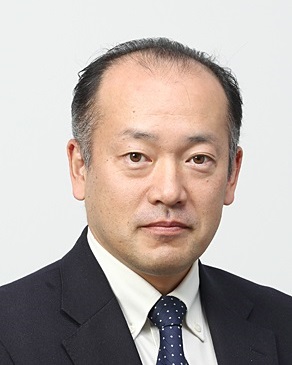}}]{Tomoo Ushio} (Senior Member, IEEE) received the B.S., M.S., and Ph.D. degrees in electrical engineering from Osaka University, Suita, Japan, in 1993, 1995, and 1998, respectively.

He was with the Global Hydrology and Climate Center, Huntsville, AL, USA, as a Post-Doctoral Researcher from 1998 to 2000.
In 2000, he joined the Department of Aerospace Engineering, Osaka Prefecture University, Sakai, Japan, as an Assistant Professor.
In 2006, he joined the Division of Electrical, Electronic and Information Engineering, Osaka University, as an Associate Professor, where he has been a Professor since 2019.
In 2017, he joined the Department of Aeronautics and Astronautics, Tokyo Metropolitan University, Hino, Tokyo, Japan, as a Professor.
His research specialties are radar-based remote sensing, passive and active remote sensing of atmosphere from space-borne platforms, and atmospheric electricity.
\end{IEEEbiography}

\EOD

\end{document}